# Prospects in analytical atomic spectrometry

A A Bol'shakov, A A Ganeev, V M Nemets

## Contents

**Abstract.** The trends in the development of five main branches of atomic spectrometry, *viz.*, absorption, emission, mass, fluorescence and ionisation spectrometry, are analysed. The advantages and drawbacks of various techniques in atomic spectrometry are considered. Emphasised are the applications of analytical plasma- and laser-based methods. The problems and prospects in the development in respective fields of analytical instrumentation are discussed. The bibliography includes 279 references.

## I. Introduction

Analytical atomic spectrometry embraces a multitude of techniques of elemental analysis that are based on the decomposition of samples into the state of free atoms followed by spectroscopic determination of their concentrations. One can distinguish five major branches of atomic spectrometry, *viz.*, absorption, emission, mass, fluorescence and ionisation spectrometry. The first three are most popular and versatile; atomic mass spectrometry is the most sensitive. Other methods used in elemental analysis (X-ray fluorescence, nuclear magnetic resonance, Auger electron spectrometry, neutron activation analysis, *etc.*), which are not based on the analyte atomisation and unrelated to the spectroscopy of atomic vapour, are not considered in this review.

The main goals of analytical atomic spectrometry are to attain the lowest limits of detection (down to single atoms) and the broadest dynamic range (ideally, from single atoms to 100% content), suppress the matrix effect, eliminate spectral interferences, minimise the time and cost required for sample preparation, and pass from multistage to direct methods. Fundamental studies on the development of methods for absolute and standardless analyses, aimed at avoiding calibration by standards, are in progress. Ever increasing is the role of the software development, which in certain cases facilitates substantially the deconvolution of complex background components, allows digital processing (averaging) of noise and enhances the analysis accuracy due to the use of correlation models and neural network algorithms.

The development of analytical spectrometry and detection techniques is stimulated by the diverse and increasing demands in industry, medicine, science, environmental control, forensic analysis, *etc.* The development of portable analysers for the determination of elements in different media at the immediate point of sampling, which eliminates the stages of collecting, transportation and storage of samples, is one of the most important directions.

It should be noted that the development of atomic spectrometry slowed down in recent years; particularly, a trend towards a decreasing number of scientific publications occurred. Such a deceleration could be due to the fact that the overwhelming majority of issues in elemental analysis can successfully be solved using the available techniques and instrumentation.

The above statements may be supplemented by several general remarks.

1. According to the amount of scientific publications and conference presentations, the inductively coupled plasma emission and mass spectrometers (ICP-AES and ICP-MS) are the most popular. This is because of the fact that the inductively coupled plasma in argon heated to $6000-9000$ K at the electron density of $\sim 10^{15}$ cm$^{-3}$ represents a virtually ideally linear atomiser and ioniser (*i.e.*, virtually independent of the sample matrix and providing the linear relationship between the analyte concentration and the analytical signal).

2. In practice, any sensitivity required for the majority of applications can be provided by commercial instruments, in particular, ICP-MS (high resolution double-focusing ICP-mass spectrometers are used in especially difficult cases).

3. Atomic absorption, arc and spark emission spectrometers remain a less expensive alternative to IPC-MS and IPC-AES instruments. Moreover, in practical applications, atomic absorption spectrometers have held the leading position for many decades. Current proliferation of ICP-AES spectrometers approaches that of the former.

4. The throughput of analysers has reached a sufficiently high level, which allows thousands of routine analyses to be performed daily. The main current tendency is to provide faster and more efficient performance using modern analytical instruments, thus spending less time and efforts.

5. Further development is aimed at improving sample introduction and simplifying the sample preparation, reducing the total

**A A Bol'shakov, A A Ganeev, V M Nemets** St Petersburg State University, ul. Ul'yanovskaya 1, 198504 St Petersburg, Russian Federation.
Fax (7-812) 428 72 00, tel. (7-812) 428 71 62, e-mail: alexandb@mail.ru (A A Bol'shakov), tel. (7-812) 466 41 29, e-mail: ganeev@lumex.ru (A A Ganeev), tel. (7-812) 428 44 53 (V M Nemets)

**Table 1.** Main abbreviations used for particular techniques of analytical atomic spectrometry.

| Atomic Absorption Spectrometry (AAS) | Atomic Emission Spectrometry (AES) | Atomic Mass Spectrometry (MS) | Atomic Fluorescence Spectrometry (AFS) | Atomic Ionisation Spectrometry (AIS) |
|---|---|---|---|---|
| Electrothermal Atomisation AAS (ETA-AAS) | Inductively Coupled Plasma AES (ICP-AES) | Inductively Coupled Plasma MS (ICP-MS) | Inductively Coupled Plasma AFS (ICP-AFS) | Laser-Enhanced Ionisation Spectrometry (LEIS) |
| Cavity Ring-Down Spectroscopy (CRDS) | Laser Ablation ICP-AES (LA-ICP-AES) | Laser Ablation ICP-MS (LA-ICP-MS) | Laser-Excited Atomic Fluorescence Spectrometry (LEAFS) | Laser Optogalvanic Spectrometry (LOGS) |
| Intracavity Laser Spectroscopy (ICLS) | Laser-Induced Breakdown Spectrometry (LIBS) | Laser Ablation Mass Spectrometry (LA-MS) | | |
| | | Resonance Ionisation Mass Spectrometry (RIMS) | | |

cost, using new software for the automation of the analysis and data processing, making faster analysis and reducing the size of the instruments (while the earlier ICP-AES analysers could have the size of a wardrobe, the modern ICP-MS instruments measuring $110 \times 60 \times 58$ cm can be installed on a desk). An active progress is observed in the development of specialised equipment for the analysis of specific samples and determination of particular elements.

Annual reviews [1-6] on the analytical atomic spectrometry published by the *Journal of Analytical Atomic Spectrometry* (London) represent the most convenient literature source to follow the trends in this field. They summarise new information on the progress along several directions, including optical and mass spectral methods, as well as use of spectrometric analysis in environmental monitoring, industry, biology and medicine. The total number of citations in these reviews and their distribution over different analytical methods reflect the trends of modern studies in atomic spectrometry. The publications cited in these reviews may be classified (on the average, over the last 5 years) as follows: modernisation of optical methods — 83%, improvement in ICP-MS — 17%, laser-based methods (including laser ablation in ICP-MS) — 16%. The latter figure illustrates an increased interest in the development of laser-based methods of analysis (publications on routine applications were not considered here).

The American journal *Analytical Chemistry* also publishes summarising overviews [7, 8] on the atomic spectrometry every two years. A particularly interesting comparison of different spectrometric methods was carried out on a five-grade scale.[9] Several techniques at a glance were also compared.[10, 11] Moreover, several books were recently published and analytical series were supplemented with new volumes.[12-21] The information contained in these sources allows us to consider the state-of-the-art in the field of analytical atomic spectrometry.

In this paper, we present the best detection limits achieved to date in the analytical techniques under discussion. However, these lowest detection limits can hardly be realised for each element; and thus, in many real analytical situations the detection limits may exceed those cited by several orders of magnitude. Most often we cite only the latest publications, in which the reader can find references to earlier articles. The list of references by no means reflects the chronological priorities. Table 1 lists the abbreviations used for individual spectrometric techniques; however, it does not exhaust all methods discussed in this review.

## II. Atomic absorption spectrometry

While the pace of the development in modern atomic spectrometry continues to slow down (a reducing number of scientific publications, reports and presentations at exhibitions), the atomic absorption analysis continues to hold its position among the most dynamic themes, even though the history of commercial instrumentation in this field can be traced back to more than 40 years ago. Several new modifications of this method were discussed in reviews;[22, 23] first of all, it is atomic absorption spectrometry based on diode lasers (with the aim of size and cost reduction) and use of continuum spectrum sources that allow simultaneous multi-element analysis. Studies on the improvement of graphite furnaces, automated introduction of powder and solid samples, flow-injection method and the use of echelle polychromators and charge-coupled detectors (CCD) for the rapid determination are in progress.

The advantages of continuum spectrum sources (*e.g.*, a powerful xenon short-arc lamp), starting with the possibility of combining simultaneous multielement analysis with simultaneous background correction, were described.[24-26] It was shown [27] that the use of the linear photodiode arrays or two-dimensional CCD arrays as detectors facilitated reduction of the inaccuracies caused by spatial absorption inhomogeneities in electrothermal atomisers (ETA). Some of the latest developments, in particular, use of continuum spectrum sources and cross-dispersed echelle spectrometers, as well as devices for automatic introduction of powder samples, have already found their application in commercial analysers.

Several summarising papers [28, 29] were devoted to the possibilities of the use of diode lasers in AAS. The advantages of broadly tuneable vertical-cavity surface-emitting laser diodes were illustrated in Ref. 30 that also cites the original publications on different experimental applications of diode lasers as the sources of tuneable radiation for determination of absorption by Al, Ba, Ca, Cr, Cs, Cu, Gd, Hg, I, In, K, La, Li, Mn, Pb, Rb, Sm, U, Y and Zr atoms. However, a much greater number of studies on the use of diode lasers were devoted to the determination of simple molecules such as $H_2O$, $CO_2$, CO, $NH_3$, HF, HCl and *etc.* (see reviews [31-33]).[†] Simultaneous determination of several elements has been realised in multiplexing schemes with several diode lasers. It is also possible to use light emitting diodes in absorption spectroscopy.[34, 35]

The very narrow spectral width of the output emission of diode lasers (usually, $\sim 10^{-5}$ nm in a single-mode operation) is of primary importance for spectrometric purposes. The laser output radiation can be scanned over the wavelength throughout the absorption line from one wing to another. This allows one to assess the non-selective background absorption at both wings of the absorption line, thus eliminating the need for calibration with a reference beam. The signal can be detected by a photodiode without using dispersion optical devices (monochromators, spectrometers); in this case, high resolution achieved is due to the spectral narrowness of the laser output. Diode lasers can be modulated at frequencies up to 100 MHz; in combination with lock-in detection (on the fundamental or higher harmonics) of the absorption signal, this facilitates cutting off the low-frequency noise and drift effects. Moreover, the spectral resolution of diode lasers is sufficient for direct discrimination of isotopes of light (Li) and heavy (U) elements. The discrimination of intermediate isotopes is possible with the sub-Doppler saturation absorption spectroscopy.

Recently, an interest in cavity-based absorption techniques used for the many-fold increase in an effective optical path through the absorbing medium was renewed. It is known [21, 36] that in the family of such techniques the method of intracavity laser spectroscopy (ICLS) in which the sample is placed inside an active laser cavity is characterised by the lowest detection limits. However, numerous non-nlinear effects that arise in an active cavity complicate severely the practical application of this method. Both a linear response and relative simplicity are achieved with laser-pumped passive cavities. Moreover, here the diode lasers can be used for pumping. These reasons along with the need for compactness and affordability of analytical instruments have defined the rapid development of the absorption methods based on passive optical cavities.

---

† This subject was also thoroughly surveyed by T Imasaka *Talanta* **48** 305 (1999).

Here, in contrast to the ICLS technique, the sample is placed in an external (with respect to the laser) cavity with high-reflectivity mirrors. The higher the mirror quality the longer the photon lifetime in the cavity, *i.e.*, the longer is its path through the sample. Thus one can achieve pathlengths of several tens kilometres for a cavity $30-50$ cm long, which makes it possible to obtain very low detection limits. One can use, as an analytical signal, either the intensity of the radiation leaking out of the cavity or the rate of signal decay upon the abrupt interruption of laser pumping, while a laser wavelength can be scanned through the absorption line. The first version corresponds to traditional extra-cavity absorption spectroscopy and is now often called the cavity-enhanced absorption spectroscopy (CEAS). In the second version called cavity ring-down spectroscopy (CRDS), the characteristic time of the energy loss in a passive cavity is measured.

In both versions utilising passive cavities, the analytical responses are directly related to the absolute concentration of absorbing impurities in the cavity. Both techniques exhibit approximately equal high sensitivities. The use of the CRDS method requires more complicated equipment, however it facilitates the absolute calibration. The output insusceptibility to fluctuations of the excitation laser radiation is one of the major merits of CRDS. Note that in both versions one has to match the cavity spectral modes to the pumping laser mode (or modes), the piesoelectric modulation of the cavity length being often used for this purpose. This technique, although being relatively new, has been the topic of a monograph [21] and several reviews.[37–40]

The use of cavities allows one to measure absorption signals that are several orders of magnitude weaker than those in conventional absorption spectroscopy. To date, only preliminary experiments have been carried out on the combination of the plasma atomisers with the CRDS detection.[41–46] The optimum conditions for igniting the plasma used as the AAS atomiser substantially differ from those conventionally employed in the emission analysis (in the same way as the requirements imposed on plasma in ICP-AES and ICP-fluorescence spectrometry differ). At the same time, insofar as CRDS is a pulsed technique, it is reasonable to couple it with the synchronised pulsed plasma, which allows substantial reduction of the plasma background emission (this is implemented, for instance, in the plasma/laser-excited fluorescence spectrometry of gases [47]).

So far, we are aware of only one paper devoted to the electrothermal atomisation combined with CRDS detection.[48] Despite experimental difficulties associated with thermal stabilisation of the cavity while an electrothermal furnace is placed inside, it was concluded that this method holds a great promise, especially related to the possibility of using diode lasers. Detection of trace amounts of mercury vapour was an example [49, 50] showing that the sensitivity of the CRDS technique can far exceed that of the conventional atomic absorption.‡ A relatively narrow spectral range in which the dielectric mirrors retain their high efficiency is a disadvantage of cavity-based methods. This limits the possibilities of these techniques regarding the number of elements that can be determined simultaneously.

‡ This is also evident from the results of determination of gallium in flame. See M Emig, R I Billmers, K G Owens, N P Cernansky, D L Miller, F A Narducci *Appl. Spectrosc.* **56** 863 (2002).

Diode laser-based atomic absorption spectrometry has not yet left the walls of research laboratories. However, progress in the development of absorption enhancement techniques in passive cavities and the gradual increase in the number of elements detectable with diode lasers give us the grounds to hope that such commercial analysers will be available in the future. In principle, diode lasers allow determination of up to 54 elements.[29] The scope of this method can be further extended to include the elements, resonance lines of which are beyond the wavelength range accessible with diode lasers, while their lower excited states may be highly populated in the high-temperature or plasma atomisers.

Studies on the use of the hollow-cathode glow discharge plasma as a 'cold' atomiser in absorption spectrometry are in progress.[51, 52] Direct sputtering of the cathode material or the material of films coated on the cathode makes it possible in certain cases to simplify substantially the atomisation procedure and reduce the matrix effects on the results of determination of impurities owing to efficient dissociation of all sample components in the plasma. In principle, by using plasma atomisers, it is possible to simultaneously acquire the absorption and emission spectra. A miniature low-frequency $(5-20$ kHz) discharge in combination with a semiconductor laser source was proposed [53] as a prototype of absorption sensors.

Modernisation of conventional atomic absorption instrumentation leads to both the realisation of multielement analysis (by using continuum sources or sets of multielement lamps) and the increase in data acquisition rates, *i.e.*, rapid sequential determination of several elements in one sample. The latest improvements can boost the characteristics of the atomic absorption analysers with electrothermal atomisers (ETA-AAS) to approach those of the ICP-AES instruments, without compromising their simplicity and low cost. Although the dynamic range of AAS spectrometers usually does not exceed three orders of magnitude, their detection limits are low, *viz.*, down to $\sim 10^{-10}\%$ for electrothermal atomisation and $\sim 10^{-7}\%$ in flames. Using ETA-AAS, one can perform sequential determination of about 50 elements with high sensitivity.

Physico-chemical processes that occur in a graphite furnace with coatings and chemical modifiers used for the sample stabilisation, reduction of the graphite corrosion and decrease in the effect of thermochemical reactions on the graphite tube were considered.[54–56] A kinetic model of the absorption signal was developed.[57] Studies on further corrections for the non-selective background absorption [58, 59] and use of the principal component analysis [60] are in progress.

The application of AAS with gaseous sample introduction (in the case of volatile hydrides, halides, organometallic and other compounds) is being gradually expanded. The atomisation is provided by decomposition of the molecular vapours, *e.g.*, in a heated quartz tube. The merits of this method include nearly 100% efficiency of the gas-phase transport and low detection limits ($\sim 10^{-8}\%$). Determination of quite a number of elements (Ag, As, Au, Bi, Cd, Co, Cu, Ge, Ni, Pb, Sb, Se, Sn, Te, Zn, *etc.*) is possible by this method, however most often volatile hydrides are used for the determination of arsenic and selenium.[61–64] Some of commercial instruments for atomic absorption are equipped with exchangeable atomiser modules, which include flame, ETA, and also a quartz tube for the introduction of volatile compounds and mercury vapour.

The concept of absolute absorption analysis is subject to continuous development. Thus, L'vov believes [65, 66] that the possibility of the analysis without using standards for calibration might be realised in the future. For this purpose, one would need spectrally narrow and bright radiation sources, thermally stabilised atomisers that can provide complete atomisation in a defined volume, and matrix array detectors that would take into account all spatial inhomogeneities in the analysed vapour distribution. In our viewpoint, the former and latter conditions can be best met with diode lasers and two-dimensional CCD detectors.

## III. Atomic emission spectrometry

In the field of atomic emission spectrometry, the leading role is played by ICP-AES. Such instruments have been commercially available for more than 30 years. Besides using conventional inductively coupled argon plasma in atmospheric-pressure torches, the emission spectra are also excited in other kinds of plasma, including direct-current plasma, glow discharge, radio-frequency discharge, microwave-induced plasma, as well as inductively coupled discharges at reduced pressure and other plasma sources. Several fresh reviews on this subject can be found in the literature [67–71] (see also the monograph [17]). Such conventional sources as sparks and arcs are still used in practice.[72, 73] The plasma of laser-induced breakdown at the focus of the laser beam in the bulk of gas (air) or on the sample surface represents a special case.

Further progress is observed in advanced light-collection schemes in ICP-AES. An increase in the optical throughput of the instruments is achieved by using both aspherical optics and high-efficiency holographic gratings. Echelle gratings incorporated into the cross-dispersed schemes are utilised [74–76] with the aim of collecting the whole spectrum on a single CCD array detector, thus providing virtually simultaneous recording of the whole data envelope, including information required for the automatic background correction. However, due to the high density of information per pixel, the spectral resolution is compromised in such schemes; hence, in the latest versions of ICP-AES instruments the plasma emission spectrum is dispersed along the Rowland sphere segment, with 10 to 20 CCD cameras installed along the spectrum span. This results in excellent resolution and very rapid recording. Most of the ICP-AES instruments allow simultaneous determination of 75 elements with a linear dynamic range of 5 – 6 orders of magnitude.

Simultaneous recording of a whole cluster of spectral lines opens up the possibility of more advanced processing of analytical information, particularly, when several lines for each element are considered.[77] Thus, one can reduce or correct the matrix effects on the results of analysis, avoid systematic errors due to self-absorption of strong lines, take into account the temporal instability of the electronic gain factor and the baseline drift, utilise multicomponent calibration and the regression relationships. Sometimes spectrometers filled or purged with inert gases are used, which enables recording of the vacuum UV spectrum where the most sensitive lines of many elements are located and fewer spectral interferences are found.

Microwave plasma with a relatively low gas temperature and high energy of free electrons is often used for the elemental analysis of compounds separated by chromatography or capillary electrophoresis. The advantages of this kind of plasma include the high excitation capability combined with the low power and compact equipment. However, due to the low temperature of plasma, its parameters strongly depend on the amount and composition of the introduced sample eluents. The analytical applications of the microwave plasma were surveyed.[78, 79]

Interesting publications [80–82] were devoted to the combination of the graphite furnace atomisation with microwave plasma excitation of emission from the analysed vapour, wich provided high atomisation temperature within a compact instrument and made it possible to determine all elements at once in microsamples. Such plasma in a graphite furnace can also be used for the elemental analysis of compounds separated by gas chromatography.[81] The dependence of the plasma parameters on the sample matrix composition (especially those rich in easily ionised elements) cannot be totally avoided; however, this dependence is weaker than that in conventional schemes where both atomisation and excitation take place in a microwave plasma.

Aimed at miniaturisation of spectrochemical analysers, keen attention is paid currently to the development of microplasma sources and analytical techniques based on their application. Echelle [75] and other microspectrometers (with a satisfactory spectral resolution and the cigarette-pack dimensions) can be used to detect the emission. The evident advantages of such miniature analysers are the possibility of using them in field and constraint environments, the very small consumption of samples and reagents, the low cost and the possibility of forming wide sensor networks with a large number of remote detectors. Several reviews [83–85] were devoted to the recent advances in the development and application of microplasma (sources of plasma volume ranging from $10^{-6}$ to $10^{-2}$ cm$^3$ were discussed).

Rapid progress is observed in the development of methods and instrumentation in the field of laser-induced breakdown plasma (laser spark) used for the analysis of solids, liquids, gases, suspended dust and aerosols. The interest arises due to the possibility of developing universal emission analysers suitable for any kind of samples, including microscopic samples, for all elements simultaneously, with excellent spatial resolution along the surface and in depth, with no contact to the samples (remotely) and no sample preparation, that are designed for the real-time operation in a compact portable fashion. In practice, the greatest difficulties are associated with calibration ambiguities and modest detection limits ($\sim 10 - ^3$% with a relative error of 5% – 10%). In many cases, only approximate calibration could be achieved. In the last 5 years, more than 1500 publications were devoted to the development of this method. The theory and diverse analytical applications of laser-induced breakdown spectrometry (LIBS), including laser-induced breakdown in gases and on ablated surfaces, were the subject of a book [18] and several reviews.[9, 40, 86–91]

To date, the major mechanisms of the formation and development of laser-induced plasma are well understood and can be predicted based on models. For the purposes of the analysis, it is desirable that the extracted mass is independent of possible heterogeneities in the sample; hence, very short pulses should be used (to eliminate the effect of thermalisation and heat transfer across the sample). For pulses of $\sim 1$ ps and shorter, the duration of the laser field is sufficient to heat up only the electrons in the vicinity of the laser-illuminated spot on the sample surface. An explosion-type ejection of these electrons causes simultaneous destruction and ejection of the 'cold' lattice material due to the Coulomb forces. Thus, in the femtosecond mode ($\leqslant 1$ ps), the process of instantaneous vaporisation and ionisation of the sample can be greatly simplified, without inducing any transfer of heat across the sample and any shielding of the laser irradiation by the plasma plume that forms only after the end of the laser pulse. These factors improve significantly the reproducibility of the analysis.

The use of short-wavelength (ultraviolet) lasers provides more efficient and reproducible ablation and hence higher accuracy of the analysis as compared to that attainable with less intricate infrared lasers. Advantages of the ultraviolet irradiation are associated with more efficient absorption in the sample bulk (i.e., smaller losses due to reflection from the surface and less absorption in the plume above the sample surface), as well as due to the lower degree of the multiphoton ionisation as compared to the laser-induced breakdown in the infrared range. Because of the higher photon energy, the ultraviolet radiation can break lattice bonds even in a single-quantum absorption event.

Thus, the ablation by short-pulse UV lasers is a direct and the most efficient method of microextraction with the lowest probability of fractionation effects, which may cause the ratios of elements in a plume to differ from the stoichiometry of the sample. This occurs if the radiation power density is sufficient for a robust laser-induced breakdown; and this condition explicitly distinguishes the laser spark from the laser desorption mode. Although the fractionation cannot be entirely avoided in any case, it is the ablation with short-pulse UV lasers that allows one to reduce the difficulties associated with calibration, reproducibility and accuracy of the analysis.

The elimination of heat transfer and meltdown effects in the femtosecond mode becomes of particular significance for local lateral and layer-by-layer analyses where one must avoid thermal 'mixing' of neighbouring areas in the sample. Moreover, the UV focus spot can be reduced to a smaller size compared to the infrared. At the same time, use of short-wavelength lasers results in deterioration of resolution in layer-by-layer analysis (due to deeper penetration of the UV laser field into the sample).

In a laser spark, very hot plasma is formed (up to 40 000 K at the electron density up to $\sim 10^{18}$ cm$^{-3}$). Moreover, plasma in plumes extracted from entirely different samples often exhibit similar characteristics. However, despite the efforts to minimise any differences, the amount of vaporised material and the plasma parameters depend on numerous factors. Therefore, laser-induced plasma as an atomiser is inferior to analytical ICP in linearity, completeness of atomisation and versatility (*i.e.*, the degree of insusceptibility to the matrix effects). As in a conventional ICP, argon is the ideal atmosphere for the formation of a plasma plume during laser spark. However, ambient air mixed with the vaporised sample material appears to be the most common medium.

The LIBS enables analysis of substances in any state, while recording the spectra on CCD arrays makes it possible to determine all emitting elements in each pulse. Numerous examples of using laser-induced plasma in emission spectrometers range from the detection of gases and aerosols in air [92–94] to the analysis of melted minerals and steels in industry [95, 96] and even rocks on other planets. [97, 98] The recent development of miniature emitter modules of the laser-amplifier type [99] brings a hope that in future the LIBS analysers will fit into a flashlight-size module connected to the control-panel notebook by a fibreoptic cable. The examples of compact laser-induced plasma spectrometers are documented. [88, 89, 95, 100–103]

Sometimes double-pulse lasers are used for reducing the detection limits in LIBS. Ideally, the first short-duration UV laser pulse induces extraction (generates a plume), while the second longer pulse in the infrared provides additional heating of the plasma in a plume, because the IR radiation is more efficiently absorbed by the plasma (*i.e.*, heats up the free electrons to a greater extent) as compared to the UV radiation. Due to this heating, more complete atomisation in a plume is achieved. [92, 96, 104, 105] In the other similar scheme, the second heating pulse uses the laser wavelength tuned in resonance with the absorption line of the main component of the sample. [106] Thus, one can reduce the detection limits by more than an order of magnitude.

The most direct way of achieving complete atomisation of the laser extraction products is their transport with a flow of argon into a conventional ICP torch. In this fashion, laser ablation is merely used for converting the sample into the aerosol state, *i.e.*, as a sampling method, followed by the analysis using ICP-AES or ICP-MS (see reviews [40, 87, 90, 91]). This reduces the detection limits down to $\sim 10^{-7}$% at the expense of merits of compactness and viable remoteness of the analysis. The main application area of laser-ablation ICP-AES (LA-ICP-AES) is the analysis of heterogeneous samples with a micron resolution. [107]

Note that the unique nature of laser ablation with direct recording of emission spectra, *i.e.*, LIBS, is fully revealed only if a remote or local microanalysis is required, since a classical electric spark (as opposite to a laser spark) can be easily made more powerful with a multi-fold superiority in the analytical volume and, correspondingly, in sensitivity. The cost of equipment for the electric spark is also much less than that for laser ablation. Moreover, the analysis of electrically non-conducting samples with no pretreatment, as well as the analysis of aerosols, dust and ash in gases can well be performed using compact electric spark sources. [108, 109] Nonetheless, the laser sparks provide more flexible control over the discharge parameters relative to the electric sparks. The laser-induced breakdown is free of a drawback typical for the electric sparks, *viz.*, the fast uncontrollable movement (twisting) of the contracted discharge body. The higher temperature in the laser-induced plasma determines both the higher degree of atomisation and the linearity of calibration, and reduces spectral interferences.

Yet another relatively simple, rapid and inexpensive method for the analysis of surfaces and near-surface layers employs 'cool' glow discharges (usually at reduced pressure and room temperature) used for simultaneous atomisation, excitation and ionisation of samples. Both direct-current and capacitively coupled radiofrequency (RF) discharges using both hollow and planar cathodes are utilised (see reviews [69, 70]). The detection limits of hollow cathode-based analysers are usually somewhat better as compared to the planar-cathode instruments (order of $\sim 10^{-4}$%). The advantage of radiofrequency discharges is the possibility of direct sputtering (*i.e.*, plasma-induced atomisation) of electrically non-conducting samples and coatings.

Gradual sputtering of the sample surface in a 'mild' discharge and the continuous recording of the emission spectra allow one to carry out the layer-by-layer analysis with unprecedented high resolution ($\sim 1$ nm of depth). A sputtering rate can be further decreased using a pulsed discharge. It is also advantageous to use pulsed discharges for the analysis of samples especially sensitive to high temperature (*e.g.*, films and polymers). Moreover, with the use of pulsed discharges, one can sometimes obtain two-dimensional spectral images of the surface, *i.e.*, perform an instantaneous analysis of heterogeneous surfaces. Using glow discharges, it is possible to analyse both very wide surface areas and very small samples. Application of different gas mixtures in glow discharges was reviewed; [110] while analytical applications in the layer-by-layer analysis of thin films and for environmental purposes were also surveyed. [111–113]

Standardless emission analysis is possible if one has an adequate theoretical model that relates the intensities of spectral lines to the concentrations of elements in a plasma. [114] Dense plasmas such as an arc, ICP and a laser-induced breakdown can sometimes be considered within the framework of the local thermodynamic equilibrium (LTE). In the LTE model, the populations of all emitting states of atoms and ions are unambiguously related to the plasma temperature and the concentrations of elements, according to the Saha – Boltzmann law. By means of deriving the temperature from relative line intensities, one can then determine the concentrations of elements with normalisation at the sum of all concentrations set equal to unity. Such a procedure of the standardless analysis (including the effect of self-absorption of lines) was developed for the case of the laser-induced breakdown spectroscopy (see Ref. 115). [§] A similar idea was used for indirect calibration of ICP-AES in the analysis of gases. [116] Generally, the models more complicated than LTE are often used for simulating plasmas (such as microwave, glow discharge, and *etc.*).

## IV. Atomic mass spectrometry

In the mass spectrometric analysis, there is a number of different methods for ionisation of the analyte atoms. They include ionisation in plasma sources (electron impact, collisional charge exchange, plasma-chemical reactions or sputtering of secondary ions from the surface), electrothermal ionisation (heated filaments and graphite cups), ionisation of vapour in the resonance laser field, radioactive source ionisation, *etc.* The most comprehensive reviews on analytical atomic mass spectrometry can be found in books [17, 19, 20] and papers. [3, 117–120] Each issue of the *Journal of Mass Spectrometry* (UK) contains a list of recent publications in the field of mass spectrometry.

Analytical mass spectrometry based on ionisation in plasma plays the leading role in the elemental analysis due to its capability of highly sensitive, multicomponent, versatile and rapid analysis.

§ For recent publications on this topic, see I B Gornushkin, A I Kazakov, N Omenetto, B W Smith, J D Winefordner *Spectrochim. Acta, Part B* **60** 215 (2005).

To date, the potential of contemporary ICP-mass spectrometers (the first commercial instruments appeared more than 20 years ago) represents virtually the reference point for all other methods and instruments. Indeed, ICP-MS enables the simultaneous determination of several dozens of elements at trace concentrations down to $\sim 10^{-9}\%$ in solid samples and $\sim 10^{-11}\%$ in ideal aqueous solutions, within a dynamic range of up to 9 orders of magnitude.

This technique is applied most often to the analysis of environmental and geological samples, and high-purity substances. However, not only the ICP but also other kinds of plasma are widely used for ionising atoms, particularly, in glow and corona discharges, and also in laser-induced plasma (laser ablation and laser desorption from the surface). Inductively coupled and corona discharges are usually ignited at atmospheric pressure, whereas glow discharge is ignited at low and reduced pressures. Corona discharge is utilised in the mass-spectrometric analysis of gases.[121, 122] Electric spark in vacuum is also successfully used as the atomiser/ioniser of conducting and non-conducting samples (the latter are pre-ground and pressed into graphite electrodes).[123]

In the majority of cases, the IPC-MS instruments incorporate quadrupole mass spectrometers (they are usually more expensive than the AAS and ICP-AES analysers, but less expensive than other kinds of mass spectrometers). One of the main problems related to quadrupole-based ICP-mass spectrometers concerns the isobaric interferences, i.e., overlaps of isotopes of different elements with equal charge-to-mass ratios. Moreover, in the argon plasma, in addition to the proper atomic ions, some parasitic molecular ions are also formed such as $Ar_2^+$, $ArO^+$, $ArH^+$, $ArOH^+$, $ArN^+$ and other, as well as multicharge ions, which introduce additional interferences. This can be alleviated if the high-resolution magnetic sector mass spectrometers in a double-focusing mode (with crossed electrostatic and magnetic fields) are used.[124–127] High-resolution instruments are very expensive but highly sensitive and facilitate recording of extremely small differences in the isobaric masses of elements.

An attractive innovation for minimising the isobaric interferences is related to the development of the RF multipole reaction cells that can be installed in conventional quadrupole mass spectrometers.[128–131] The cell is filled with a reactive gas (e.g., ammonia, oxygen, or methane) under reduced pressure, and the field-driven analyte ions drift towards the main mass analyser (quadrupole) and the detector. In the simplest case, the buffer gas molecules collide with the parasitic molecular ions produced in an ICP and break them apart into the atomic form. However, transforming particular parasitic ions into the molecular ions with a different mass in ion–molecule charge exchange reactions is a much more efficient way of eliminating specific interferences. Quadrupole, hexapole and octopole systems are used as RF multipole cells.[131] Sometimes RF ion traps are used for this purpose. The best performance was observed for quadrupole dynamic reaction cells, which can be tuned synchronously with the main mass analyser (i.e., in resonance with the analyte mass), thus cutting off possible secondary interferences in the upper and lower mass ranges. As a result, the common quadrupole mass spectrometers equipped with RF multipole reaction cells are presently a sound alternative to the very expensive high-resolution magnetic sector mass spectrometers.

Since the quadrupole mass spectrometers are the scanning devices in their principle of operation, the speed (throughput) of detection is low in the multielement analysis, because the acquisition time required for determining each element in the vicinity of its detection limit is about $\sim 10$ s. As a result, further efforts take place in the development of magnetic sector mass spectrometers with multichannel/multicollector detectors for simultaneous detection of all ions arriving to the focal surface.[125, 132–134] For the similar purposes, the time-of-flight mass spectrometers that can simultaneously record the whole spectrum of elements in each ion extraction pulse, are being used.[135–138] There are commercial models of the ICP-time-of-flight mass spectrometers.

In contrast to magnetic sector multichannel (multicollector) mass spectrometers, the time-of-flight mass spectrometers have long been used in analytical studies and are available in various models. However, potential advantages of magnetic sector mass spectrometers are related to the fact that they are continuous, and consequently, superior in the integration time, accuracy, stability and noise averaging possibilities. Time-of-flight mass spectrometers can determine ions in a broad range of masses simultaneously; moreover, they are less bulky and less expensive. A certain method of simultaneous recording is specifically required in those cases where fast or transient processes are studied, e.g., in chromatographic or electrophoretic separation, in flow-injection analysis, electrothermal vaporisation, laser ablation or laser desorption.

A combination of ICP-MS with the direct sampling based on laser ablation (LA-ICP-MS) facilitates circumventing the complex and lengthy sample preparation and atomisation procedures. The reviews on this subject have been mentioned above.[87, 90, 91, 120] For the accurate trace determination by ICP-MS at very low detection limits, the sterile direct laser sampling plays an important role because omitting the sample preparation stage reduces sharply the probability of accidental contamination during sample preparation to the analysis. At the same time, undesirable fractionation of elements during the aerosol transport along the analyser ducts does sometimes occur in LA-ICP-MS.[139] The calibration reproducibility and accuracy may also become difficult to assure. In those cases where the acquisition of spectra from each laser pulse is essential, the ICP-sector mass spectrometers with detector arrays[140] or ICP-time-of-flight mass spectrometers[141] are used.

The overall dimensions of the instrument can be substantially reduced by using the laser-induced plasma directly as an atomiser/ioniser in mass spectrometry.[142] Usually, the time-of-flight mass spectrometers are employed in laser ablation mass spectrometry (LA-MS). In this case, the pulsed nature of the laser ablation almost perfectly matches the pulsed extraction of ions.[143–146] Such instruments surpass the laser-induced breakdown emission analysers in sensitivity and could have been quite competitive in regard of compactness[147] if continuous evacuation was not necessary. Using ablation at the resonance wavelength, one can further reduce the detection limits.[148, 149] Laser irradiation at lower power does not cause any crater-like destruction of the analysed surface, thus atomic and molecular ions can be detached from it in a desorption mode. Measurement of these ions with time-of-flight mass spectrometers enables very careful analysis of the art, jewelry and documents for forensic examination.[150]

Thin films, coatings and other surfaces can be sputtered layer-by-layer in glow discharges with excellent depth-profiling resolution (but measurements are in emission most often). Using a mass spectrometer as the detector,[69, 117] one can reduce the detection limits by several orders of magnitude (down to $\sim 10^{-7}\%$ for single element detection). The uncertainties of quantitative calibration are lower[151, 152] in this case than in LA-ICP-MS. This method was successfully used for analysing the environmental samples and nuclear waste.[113] In the ICP-MS spectrometers manufactured by Finnigan MAT, the ICP module was exchangeable with a direct-current glow discharge module. Recently, high-resolution mass spectrometers with ionisation in the direct-current glow discharge and the radiofrequency discharge have been introduced by VG Elemental. Despite their high cost, they have already found various applications ranging from analysis of high-purity metals to analysis of powders and dielectrics. The analysers that include both a pulsed glow discharge and a time-of-flight mass spectrometer have recently started to be produced in Russia.[153]

Mass spectrometry can also be integrated with other plasma sources, such as microwave plasma[154, 155] and capacitively coupled RF plasma in a graphite furnace.[156] However, these sources are used much more often in emission spectrometry. Their advantage is the possibility of small sample volume intro-duction, including a continuous (flow-injection) mode. Microwave plasma enables the analysis at the temperature near ambient, whereas the discharge in a graphite furnace makes it possible to maintain high atomisation temperature with the independently regulated efficiency of ionisation in plasma. A mass-spectrometric detector facilitates the reduction of the detection limits relative to the corresponding emission detection; however, some setbacks in the design still remain unsolved. Application of such instruments is justified in those infrequent cases where the analytical problem cannot be solved using common ICP-MS analysers (e.g., due to the specific spectral interferences).

The capability of determining different isotopes and their ratios is also among the advantages of mass spectrometry. To date, the thermal ionisation mass spectrometry continues to be the most precise technique for measuring isotope ratios (for a comparison with other techniques, see Refs 157 and 158). However, this technique requires complicated sample preparation and is almost inapplicable to both refractory substances and elements with high ionisation potentials. The precise data on the isotopic composition are necessary for studying the synthesis, evolution or the absolute age of samples (geological, organic, biological, etc.). Conventional methods of determination of stable isotopes in gaseous samples (and any other samples that can be converted into the gaseous phase) using magnetic sector mass spectrometers were surveyed [159] with a focus in climatology.

A unique possibility of minimising the isobaric interferences in mass spectrometry is offered by selective ionisation in the resonant laser radiation (resonance ionisation mass spectrometry, RIMS). A laser (or several lasers) is used for the optical excitation and selective stepwise transfer of the analyte atoms into the ionised or auto-ionised states. Sometimes the atomiser is placed in the electric field to reduce the effective ionisation potential of the atom. Examples of the atomisers include a graphite furnace,[160] or low-power laser ablation [149, 161] or laser desorption [162] at low power to minimise the non-selective ionisation. Use of tuneable solid-state lasers and particularly diode lasers holds a promise for this technique.[163] The main application field is the determination of stable and long-lived isotopes as well as radioactive contamination. The detection limits can be below a femtogram at a $\sim 10^{-11}\%$ relative content of the rare compared to the abundant isotopes. This method has already been used for determination of more than 40 elements; in some cases multielemental isotope analysis was performed.[164] Laser ionisation schemes for many elements were described.[165] Recent information on RIMS has been reported.[166, 167]

High-energy accelerator mass spectrometry, which utilises tandem linear particle accelerators, is aimed at the determination of cosmogenic and other very rare isotopes ($^3$H, $^{10}$Be, $^{14}$C, $^{26}$Al, $^{32}$Si, $^{36}$Cl, $^{39}$Ar, $^{41}$Ca, $^{55}$Fe, $^{59}$Ni, $^{81}$Kr, $^{99}$Tc, $^{129}$I, $^{182}$Hf, $^{244}$Pu).[168, 169] This method can be used for the determination of very low amounts of particular isotopes (at isotope ratios down to $\sim 10^{-14}\%$). Such a high sensitivity is determined by both several selection stages discriminating the ion mass and charge and selection in the kinetic energy loss per track unit in the gaseous or semiconductor detector. The specific energy loss depends on the nucleus charge (not the ion charge) that facilitates effective elimination of the isobaric interferences, however such detectors can determine only particles with high velocities. Sputtering of samples is usually performed using a beam of $Cs^+$ ions in vacuum. This method is applied in archaeology, dendro- and geochronology, climatology, medicine and forensic examination.[168–172] The equipment usually occupies an area of several tens square metres; however new instruments with dimensions of only $2.25 \times 1.25$ m$^2$ are already under development.

The theory, instrumentation, and applications of ion mobility spectrometry are being actively developed (see a review [173]). This technique differs from the classical mass spectrometry in a way that the ions are separated by their charge-to-mass ratio while drifting in the gas medium driven by the electric field, rather than in vacuum. The main advantage is that there is no need for evacuation and, correspondingly, the cost is lower and miniaturisation of such high-sensitivity instruments is possible. The ion mobility analysers have been applied so far only for the gas analysis (detection of contaminations and toxic compounds in air or trace impurities in technological gases).[174–177] They will possibly find broader applications with the progress in this field because of the obvious advantages of combining such detectors with non-vacuum ionisation sources (e.g., laser ablation and desorption, discharges at atmospheric and slightly reduced pressures, including microplasma discharges). An increasing body of information on applications of RF multipole reaction cells (based on drift and ion reactions in the gas medium) in mass spectrometry will also favour development in ion mobility spectrometry.

For quantitative calibration, especially in ultra-trace mass spectrometric analysis where the standards cannot be prepared, the method of isotope dilution is often used, e.g., in ICP-MS.[178, 179] An increase in accuracy of the semiquantitative analysis in ICP-MS can be achieved by simulation of the plasma-chemical processes.[180] Worth to note are the studies devoted to the calculation of the relative sensitivity [181] and the standardless analysis [182] in ICP-MS and LA-MS.[183] The possibility of simultaneous acquisition of spectra from atoms, molecules and fragments represents another advantage of mass-spectrometry. Moreover, this technique can be hyphenated to chromatography or other methods of preliminary separation of chemical compounds (see a review [184]).

# V. Atomic fluorescence spectrometry

At present, the atomic fluorescence spectrometry (AFS) plays a relatively minor role in elemental analysis. An attractive feature of the fluorescence techniques is the fundamental possibility of combining the advantages of absorption and emission spectrometry. Theoretically, the analytical signal is proportional to the concentration of atoms in the (most populated) ground state as in the case of absorption; but it is being measured on a 'zero' or weak background as in emission spectrometry. The selectivity of atomic fluorescence is very high because both mechanisms of absorbing and emitting of photons are discrete quantum processes. Excitation of the selected atoms only on their characteristic spectral lines and shielding the detector from direct rays of the light source enable measurement of fluorescence without recourse to any optical dispersion devices. Moreover, one can modulate the excitation source for better subtraction of the stray light background.

These advantages determine the application area of AFS. The development occurs in two major directions. Where high selectivity and high sensitivity are required that cannot be reached by other methods, one can use the laser-excited fluorescence (pulsed most often) with detection by means of optical and electronic schemes that are optimised for subtracting the non-selective background, reducing the spectral interferences and minimising the noises. In the other cases where moderate sensitivity is sufficient, while a simple and rapid analysis procedure and inexpensive equipment are desirable, then simple optical filters (in place of dispersion spectrometers) and traditional light sources, e.g., hollow cathode lamps or RF electrodeless lamps, are used. Naturally, weak sources and thus unsaturated fluorescence bring about a number of drawbacks including the non-selective absorption of radiation in the atomiser.

In line with the development in chemistry of stable and unstable volatile compounds, there is an increasing interest in the simplest version of AFS that involves continuous sample introduction in the form of vapour (e.g., mercury [185–187]) or volatile hydrides [188–193] with their low-temperature atomisation. High selectivity of this technique facilitates analysis of samples of a very complex chemical composition with high sensitivity. Modern commercial atomic fluorescence analysers [194] employ hollow-cathode lamps, graphite atomisers, and continuous flow-injection introduction of samples in the form of volatile hydride compounds. The detection limits are often below $10^{-7}$%. Note that it is the detection by fluorescence or absorption techniques rather than by emission that is more suitable for these compounds, because introduction of large amounts of volatile compounds into a plasma almost always causes instabilities in the discharge.

Commercial ICP-AFS instruments made earlier in USA were equipped with hollow cathode lamps and conventional nebulisers for sample introduction. These instruments were used for the determination of certain metals with weak emission spectra but easily excitable fluorescence spectra. The plasma was ignited in a relatively 'cool' regime in order to merely atomise the sample without generating too high emission background. Further development of this method is associated [195] with the use of new light sources of high spectral brightness, namely, diode lasers or narrow-band light-emitting diodes. These sources can also improve the characteristics of the simplest fluorescence analysers that involve introduction of samples in the form of vapour or volatile compounds.

One of the main advantages of the laser-excited fluorescence is that it makes possible the saturation (bleaching) of optical transitions. Thus, it is possible to obtain maximum analytical signals and simultaneously reduce both the effects of fluorescence quenching and its dependence on fluctuations of the excitation radiation. Unfortunately, during the past 25 years, laser-excited atomic fluorescence was developing very slowly (see a review [196]) owing to the slow progress in the development of tuneable laser technologies. The real-time switching of dye lasers from one element to another is virtually impossible. It is the single-element nature of this technique that presents the main obstacle for its analytical application, although more than 40 elements were determined one by one using AFS with dye lasers.

The dynamic range of laser-excited atomic fluorescence spectrometry (LEAFS) can reach 7 orders of magnitude (much broader than in AAS). Moreover, LEAFS is characterised by better linearity compared to AAS. The lowest detection limits in LEAFS are of the order of $\sim 10^{-11}$% ($\sim 0.1$ fg) in the analysis of ideal aqueous solutions using an electrothermal atomiser.[196] The best results for low-volatility elements (Eu, Y, Tl) were obtained using atomisation in a glow discharge. Use of flame, ICP, laser spark and other plasma atomisers has been studied. In the analysis of gases, the plasma was used for the pre-population of the metastable states, from which further excitation of laser-induced fluorescence was performed.[47]

About ten years ago, new solid-state laser sources appeared;[197–200] they are the optical parametric oscillators (OPO) with tuneable output radiation, which can be scanned over the whole spectrum from 220 to 2000 nm in $\sim 20$ min (previously, such sources were called parametric light generators). Using the modern software, it is possible to scan discretely the OPO output radiation, *viz.*, by jumping from one characteristic line to another, thus facilitating reduction in the scanning time (similarly to the newest analytical mass spectrometers). Obviously, OPO open up the prospects for the development of multi-element LEAFS. However, so far only few studies devoted to simultaneous detection of several elements using the OPO-excited fluorescence technique have been accomplished.[201, 202]

In our viewpoint, the commercial LEAFS instruments based on OPO can be developed in the future, despite their high cost at present. An ICP-AFS spectrometer based on OPO would be comparable in cost with a high-resolution ICP-mass spectrometer. However, despite the virtually equally high sensitivity of mass spectrometry and laser-excited AFS, the selectivity of the latter technique is higher relative to high-resolution mass spectrometry and much higher relative to either quadrupole or time-of-flight mass spectrometry. Furthermore, theoretically, the sensitivity of AFS should be higher than that of mass spectrometry, because each atom can produce only one ion but can emit up to $\sim 10^8$ photons per second. The throughput of mass spectrometers is usually $\sim 10^{-6}$ of the ion flux, whereas the typical efficiency of collection and registration of photons is of the order of $\sim 10^{-4}$.

Perhaps, the commercialisation of LEAFS analysers will be aided by the development of multifunctional laser spectrometers with modular architecture aimed at enabling application of several analytical techniques within a single instrument.

New potentials in LEAFS are opening up with the development of magneto-optical traps (laser traps) for neutral atoms.[203] Single atoms including their particular isotopes can be captured in a laser trap, and the resonance fluorescence of each individual atom can be reliably detected. Using this method, rare isotopes $^{81}$Kr and $^{85}$Kr with the natural abundance of the order of $10^{-11}$% and $10^{-9}$%, respectively, were determined.[204] The detection limit of the $^{41}$Ca isotope in biological samples atomised in a furnace was below $10^{-7}$% with respect to the abundant $^{40}$Ca isotope.[205] So far, such low amounts of rare isotopes could only be detected using high-energy accelerator mass spectrometry. Certainly, the LEAFS instrumentation based on magneto-optical traps is less expensive and less massive. Laser traps have already been successfully used for confinement of alkali and alkali-earth atoms, as well as noble gas atoms (pre-excited in plasma into the metastable states for laser manipulation).

It is hoped that further development of atomic absorption techniques based on diode lasers will also be beneficial for their application in AFS. The high spectral brightness of the diode lasers (due to the very spectrally narrow output emission) at their typical optical power of $\sim 1$ mW with a beam focused into a zone of a diameter of $\sim 1$ mm allows one to achieve complete saturation of fluorescence on strong transitions, if the quenching is not considerable. Multielement analysis can be achieved (as in AAS) by using diode laser arrays incorporated into multiplexing schemes. However, so far AFS with excitation by diode lasers has been used only in a single-element fashion.[206–212]

Similar to the absorption spectrometry, the saturated (laser-excited) AFS provides a theoretical possibility of the absolute analysis. This possibility arises from both the simple theoretical dependence of the saturated fluorescence intensity on the analyte atom concentration and the reliable knowledge of the photon detection efficiency achievable in practice. In the case of AFS, the conditions originally formulated by L'vov [65, 66] as necessary for the absolute absorption analysis, are supplemented here by an additional requirement for the sufficient power of the output radiation of the excitation laser. The measurement of the non-selective background is made by detuning the laser from the absorption line. If the power of the excitation source is far from saturating, the fluorescence intensity usually depends on many unaccountable factors, which necessitatees the use of standards to implement the calibration for the quantitative analysis.

Note that at this stage the practical applications of LEAFS are largely limited by the tasks that require the highly sensitive and selective analysis. The preparation, storage and use of standard solutions for calibrating the instruments in the range of extremely low concentrations of elements pose serious problems and require certain measures to be undertaken due to the high probability of accidental contamination at any step of calibration. Hence, the development of methods for the absolute (even if semi-quantitative) analysis has gained crucial importance. The idea of the absolute AFS analysis appears even more attractive, since this method usually demonstrates a precisely linear dependence of the analytical signal on the analyte concentration.

## VI. Atomic ionisation spectrometry

The atomic ionisation analysis is based on selective stepwise ionisation (laser-induced in most cases) of atoms. It resembles in essence the RIMS technique. However, the intensity of the current formed by the electron – ion pairs upon ionisation is measured in this method rather than the ion masses. One such technique called resonance ionisation spectrometry implies usually two- or three-step laser-induced photoionisation of atoms, most often in vacuum as in RIMS. The other more popular version of the method is the laser-enhanced ionisation (LEI) spectrometry in flames at atmospheric pressure, where the analyte atoms are selectively excited by the laser irradiation into the highly excited states, thus enhancing (by up to 8 orders of magnitude) the collisional thermal ionisation. Optogalvanic spectrometry is also based on a similar scheme, which employs the plasma discharge at reduced pressures for cathodic sputtering of the samples followed by laser-induced excitation of atoms and registration of the change in the current owing to the change in the ionisation rate, that is largely of the non-thermal origin in this case.

The main advantage of all forms of the atomic ionisation spectrometry is the high sensitivity achievable with relatively simple instrumentation, especially if the diode lasers are used. The efficiency of collection and registration of single electrons in both vacuum and weakly ionised flames (the electron density in flames is usually $< 10^9$ cm$^{-3}$) can, in principle, be brought close to 100%. High-stability hollow-cathode lamps characterised by fluctuations of plasma current only at the shot-noise level, facilitate reliable measurements of the very weak optogalvanic signals ($\sim 1$ nA). The selectivity in this method with simple recording of current is lower than in more sophisticated mass spectrometric or laser-excited fluorescence detection. The main disadvantage of the method is a strong matrix dependence (especially in the presence of easily ionised elements that not only increase the background noise but can also change the electric field between the electrodes owing to the space charge formation). The interferences due to the parasitic multiphoton ionisation of the flame-generated molecules and radicals such as NO, CO, $O_2$, etc. are also encountered in laser-enhanced ionisation in flames.

Laser-enhanced ionisation spectrometry (LEIS) is the leading technique among other techniques of the atomic ionisation analysis, although its development is also impeded because the suitable tuneable lasers are still unavailable. LEIS provides low detection limits of $\sim 10^{-10}\%$ ($\sim 1$ fg) at very small sample amounts (less than 1 pg) and with a fairly simple design. Further increase in sensitivity can be achieved with avalanche ionisation, i.e., with the $10^3 - 10^5$-fold multiplied current of the primary electron–ion pairs.[213, 214] Application of nanosecond lasers and nanosecond-resolution registration electronics in some cases enables separating the fast response due to the parasitic molecular ionisation from the slower collisional ionisation response in LEIS at reduced pressures.[215] The parasitic ionisation can also be minimised by shielding the flame from air with an outer concentric argon flow.[216, 217]

For the purpose of sample introduction into the flame and more complete atomisation in LEIS, one can use either electrothermal vaporisation[218] or transformation of samples into volatile compounds,[217] or flow-injection introduction with a possibility of sorption pre-concentration.[219] The advantage of the microsample analysis in LEIS is most evident when it follows either chromatographic separation[220] or laser ablation sampling into the flame.[216, 221–223] As the laser-induced plasma creates high temperature and high densities of charged particles, one may opt to separate the sampling from the signal registration either temporally[221, 222] or spatially.[216, 223] In the latter case, the product of laser ablation (aerosol) is carried by the argon flow into the miniature flame where the LEIS signal is generated. This concept is similar to LA-ICP-MS and LA-ICP-AES. Use of a miniburner reduces the effect of aerosol dilution during the analysis. Such schemes open up the possibilities of detecting trace impurities in each laser ablation pulse (15 atoms per pulse[214]). Moreover, it is possible to simultaneously determine several elements (Pb, In) in samples.[223]

Electrothermal atomisation in an inert atmosphere of a graphite furnace can become an alternative to the use of flames in LEIS. This method provides very low detection limits for noble metals.[224] A direct comparison of the detection limits in LEIS and LEAFS under the same conditions indicates the latter to be more sensitive,[217, 225] if the optical pumping power of the laser is sufficient for the fluorescence saturation. However, under the conditions most favourable for LEIS (collision frequency $> 10^9$ s$^{-1}$), the fluorescence quantum yield is small due to the quenching processes. In the case of unsaturated transitions, LEAFS is inferior to LEIS.[220] It is fundamentally difficult to achieve complete saturation of all optical steps in LEIS, because transition probabilities to reach highly excited states are usually low.

It is interesting to compare the detection limits of LEIS and purely optical laser-induced resonance photoionisation in metal vapour at atmospheric pressure.[224] Thus for silver atoms, the two-step resonance photoionisation through the auto-ionising state provides a better detection limit ($2 \times 10^{-11}\%$) compared to the two-laser LEIS ($2 \times 10^{-10}\%$). Generally, a success of any of these schemes depends on the relationship between the efficiency of collisional ionisation and the rate of optical pumping into auto-ionising states. However, an increase in the ionisation rate provided by two- and three-step schemes, is usually several orders of magnitude greater than in a single laser excitation step. A complex laser spectrometer for the laser-enhanced ionisation in both vacuum and flame was developed for the analytical research purposes.[226, 227] Recent developments associated with the LEIS technique were discussed in a review;[228] the earlier records were systematised in monographs[36, 229] and a review.[40]

Laser optogalvanic spectrometry (LOGS) may be appealing due to its simple electric gas-discharge assembly used simultaneously for both sample atomisation and photoelectron detection. The literature records on optogalvanic detection of overall 40 elements were summarised[230, 231] (data on indium are also documented[232]). Uncomplicated design can apparently be achieved only with diode lasers,[233–235] the advantage of which is not only their miniature size but also their capability to resolve the isotopic spectra. A narrow single-mode spectrum generated by a diode laser enhances the selectivity of analysis. To our knowledge, the methods for determination of mercury[236] and uranium isotopes[233, 234] are the only real analytical applications that have been developed so far.

A prototype of the portable analyser for uranium isotopes (most important element in the nuclear power industry) based on LOGS with a modulated diode laser and lock-in detection was described.[233] This device enabled determination of isotopes $^{235}U$, $^{236}U$ and $^{238}U$ in the enriched and depleted samples of uranium oxides and fluorides. Such a method under laboratory conditions[234] enables measurements of the isotope ratio $^{235}U/^{238}U$ down to 0.02% with the detection limit in atomic number density $\sim 10^7$ cm$^{-3}$ in the analytical zone. This sensitivity is comparable to that provided by AAS with the same modulation of the diode laser wavelength.[234] A similar technique may be used for the determination of many other isotopes of actinides and lanthanides.

The LOGS technique propped with the RF electrodeless discharges[231] may eliminate introduction of the impurities by electrodes and, in principle, makes it possible to analyse non-conducting samples directly. In this case, the laser-induced changes in the plasma impedance are measured using a separate resonant induction coil. The optogalvanic spectrometry usually utilises 'cold' plasmas with the temperature of about 300 K. This is convenient, e.g., for the analysis of gases,[231, 235] but problematic for atomising aerosols. A comparison of the analytical merits of LOGS and LEIS for the detection of metals in smoke particles[237] revealed the undeniable advantages of LEIS regarding the completeness of atomisation at the flame temperature of $\sim 2500$ K.

At the same time, the characteristic noise in LEIS is determined by fluctuations of thermal (and turbulent) velocities of charged particles in flame. Thus, in principle, the noise can be reduced by decreasing the temperature. On the contrary, in LOGS, a number of atoms in the analytical zone can be increased by increasing the pressure. For this purpose, it was proposed[235] to use a non-sustained discharge at atmospheric pressure and room temperature with the electron density $\sim 10^{12}$ cm$^{-3}$ below a threshold of the self-sustained gas discharge. The electric current noise in this arrangement is reduced to $\sim 0.1$ pA ($\Delta I/I \approx 10^{-8}$). The detection limit for the excited hydrogen atoms H* in neon was estimated from an optogalvanic response to the modulated irradiation by a diode laser to be $\sim 10^{-12}\%$ (at the 656-nm line). For other elements, the sensitivity may be even higher if the laser excitation scheme provides a chance to climb closer to the ionisation potential.

A problem of the space charge formation is also largely curbed, because only a weak electric field ($\sim 100$ V cm$^{-1}$) preserving the plasma parameters unperturbed, is used for the proper electron collection, whereas in LEIS the fields of $\sim 1000$ V cm$^{-1}$ are commonly used. Note that the conditions necessary for this concept may be realised in a late afterglow of the laser-induced plasma.

Generally, the atomic ionisation spectrometry is yet of little use in practice but continues to be developed and has its potentials. As with other laser-based techniques, a hindering factor relates to a slow progress in the development of tuneable lasers that cannot yet support in practice the multielement analysis. High susceptibility of the ionisation degree to the matrix composition presents another serious drawback of this technique. A fundamental possibility of the absolute analysis is one of the potentially important merits of the atomic ionisation technique. An absolute number of the collected electrons (charge) is equal to the absolute number of the analyte atoms, the 100% ionisation of which can be easily achieved in the analytical volume limited by the intersection area of the laser beams, while a correction for the non-selective background is obtained by tuning the laser over the full profile of the analytical absorption line. It is necessary to know either the atomised mass or the degree of sample vaporisation, as well as the dynamics of vapour transport. These parameters can be estimated using mathematical models.

## VII. Sample preparation and introduction, atomisation and data processing

Here we cite only a small part of the latest reviews among the vast number of publications on the modernisation, automation and miniaturisation of the instrumentation for sample preparation,[238–241] including those dealing with the programmed microwave reaction systems,[240, 241] as well as with further integration of the sample preparation and introduction stages. In particular, there are many different methods of the physicochemical separation, pre-concentration and extraction (e.g., see a review[242] on the pre-concentration of microcomponents). Many studies were devoted to coupling of the chromatographic separation with continuous atomic spectrometric analysis of the eluates.[184, 243–247] Recently, the method of continuous flow-injection sample introduction became very popular.[248–251] The methods of sample introduction in the form of vapour of volatile compounds find an increasing number of application.[252–256]

The methods of electrothermal vaporisation[257–259] and their combinations with chemical transformations of the samples into volatile compounds for flow-injection introduction[260, 261] are also subject to further development. The possibilities[261] of complete automation of the whole process of pre-concentration, separation, vaporisation, introduction and atomisation of samples including microwave-assisted decomposition of samples, and a possibility of the direct introduction of large amounts of undiluted organic solvents, viscous liquids as emulsions and solid insoluble samples as slurries are considered. The reviews[262, 263] devoted to the simple tungsten and molybdenum electrothermal atomisers were also published. Great attention is being paid to further research on improvement of nebulisers for introduction of the analysed solutions into inductively coupled and microwave-induced plasmas.[264, 265] Many studies in elemental analysis were devoted to introduction and vaporisation of the dissolved samples using high-voltage electrohydrodynamic micronebulisers[266–269] (originally this method was named the 'Extraction of Dissolved Ions at Atmospheric Pressure' but now it is most often called electrospray).

Many studies ponder about the software issues on automation of analytical instrumentation and about new mathematical methods for data processing. The journal Analytical Chemistry publishes detailed reviews on this subject every two years (for the latest review, see Ref. 270). Most popular approaches used in the modern spectroscopy were systematised.[271, 272] A substantial improvement in both reproducibility and correctness of the analytical results could be achieved using the multidimensional matrix and cluster methods of statistical data treatment. Such methods allow one to recognise and take into account unknown active factors (e.g., the presence of hidden impurities, drift, etc.) without carrying out frequent calibrations of the equipment or to restrict oneself to only one 'permanent' calibration. In some cases, it is possible to transfer the calibration from one spectrometer to the whole family of this kind of spectrometers.[273] As mentioned above, the mathematical simulation also opens up possibilities for the absolute (standardless) analysis.

A progress in the computer-assisted methods of operation, self-diagnosis and self-control in analytical instrumentation, as well as the comprehensive processing of analytical data, is largely due to the impressive development in the neighbouring fields. In future, one can expect application of the methods of digital image recognition, machine self-learning, digital filtration of noises and processing of the hyperspectral spatial images. The computational methods using genetic algorithms and artificial neural networks will certainly find their application for spectrometric purposes.[274] There is a wealth of prospects for further development in these directions but even today the software and continuous automatic self-diagnosis enable the analytical instruments to 'advise' the operator the ways for resolving problems or identifying the possible errors.

## VIII. Conclusion

At present, AAS, ICP-AES and ICP-MS are the undeniable leaders among a multitude of the techniques in the analytical atomic spectrometry. During several past decades, the inductively coupled plasma displaced other sources of excitation and ionisation of atoms in the emission analysis and mass-spectrometry. Among these universal multielement techniques, the ICP-MS exhibits the best overall sensitivity. Electrothermal atomisers continue to play the key role in the absorption analysis. Currently, the laser-induced plasma is becoming the leading technique for the direct analysis of solids, while the further development evolves in the three main directions: direct laser-induced breakdown emission spectrometry, direct laser ablation mass spectrometry, and laser ablation with aerosol transport into ICP-MS.

In the atomic absorption spectrometry, one should expect the increasing use of both diode lasers and light-emitting diodes as the sources of radiation, as well as the use of optical cavities to expand the effective path of absorption. Application of the atomic fluorescence and ionisation will still be limited to unique special tasks, while further progress will depend primary on the development of the tuneable solid-state lasers. Nonetheless, in those cases where the extreme sensitivity is required for detecting single atoms and their individual isotopes, the laser-excited fluorescence and resonance ionisation techniques have already begun to compete with the accelerator mass spectrometry.

In the development of analytical instrumentation, the following most general trends may be outlined:

— combine all stages of the analysis in a single device including sample preparation (the latter remains being intricate and the least reproducible procedure);

— provide high sensitivity and low detection limits, while decreasing the sample amount and increasing the throughput of the analysis;

— develop more advanced software, maximum degree of automation, on-line capabilities of the instrument diagnostics,

switching on and off, and remote acquisition of the results through the Internet;

— design miniature (using diode lasers, microplasma, *etc.*), compact or portable analysers;

— make steps towards a modular architecture of the multi-functional spectrometers combining the capabilities of several analytical techniques in a single instrument.

The importance of the issues discussed in this paper is further supported in the most recent reviews on the subject.[275–279] There is no doubt that the analytical atomic spectrometry will continue to play an indispensable role in science, research, laboratory practice, industry and other diverse fields of human activity.

# References

1. E H Evans, J A Day, A Fisher, W J Price, C M M Smith, J F Tyson *J. Anal. At. Spectrom.* **19** 775 (2004)

2. E H Evans, J A Day, W J Price, C M M Smith, K Sutton, J F Tyson *J. Anal. At. Spectrom.* **18** 808 (2003)

3. J R Bacon, J C Greenwood, L van Vaeck, J G Williams *J. Anal. At. Spectrom.* **19** 1020 (2004)

4. S J Hill, T A Arowolo, O T Butler, J M Cook, M S Cresser, C Harrington, D L Miles *J. Anal. At. Spectrom.* **19** 301 (2004)

5. A Fisher, P S Goodall, M W Hinds, S M Nelms, D M Penny *J. Anal. At. Spectrom.* **19** 1567 (2004)

6. A Taylor, S Branch, D Halls, M Patriarca, M White *J. Anal. At. Spectrom.* **19** 505 (2004)

7. N H Bings, A Bogaerts, J A C Broekaert *Anal. Chem.* **76** 3313 (2004)

8. N H Bings, A Bogaerts, J A C Broekaert *Anal. Chem.* **74** 2691 (2002)

9. J D Winefordner, I B Gornushkin, T Correll, E Gibb, B W Smith, N Omenetto *J. Anal. At. Spectrom.* **19** 1061 (2004)

10. A A Pupyshev, A K Lutsak *Anal. Kontrol* **4** 141 (2000)

11. M A Bol'shov *Zavod. Lab.* **70** 3 (2004)

12. M Cullen (Ed.) *Atomic Spectroscopy in Elemental Analysis* (Oxford: Blackwell, Boca Raton, FL: CRC Press, 2004)

13. J Sneddon (Ed.) *Advances in Atomic Spectroscopy* Vol. 7 (Amsterdam: Elsevier, 2002)

14. A A Pupyshev, D A Danilova *Atomno-emissionnyi Spektral'nyi Analiz s Induktivno Svyazannoi Plazmoi i Tleyushchim Razryadom po Grimmu* (Inductively Coupled Plasma Atomic Emission Spectral Analysis with Grimm-Type Glow Discharge) (Ekaterinburg: GOU VPO UGTU-UPI, 2002)

15. B Welz, M Sperling *Atomic Absorption Spectrometry* (3rd Ed.) (Weinheim: Wiley-VCH, 1999)

16. J A C Broekaert *Analytical Atomic Spectrometry with Flames and Plasmas* (Weinheim: Wiley-VCH, 2002)

17. R K Marcus, J A C Broekaert (Eds) *Glow Discharge Plasmas in Analytical Spectroscopy* (Chichester: Wiley, 2003)

18. Y I Lee, K Song, J Sneddon *Laser-Induced Breakdown Spectrometry* (Huntington: Nova Science Publishers, 2000)

19. P Silvester (Ed.) *Laser Ablation-ICP-MS in the Earth Sciences* (St. John's: Mineralogical Association of Canada, 2001)

20. J R de Laeter *Applications of Inorganic Mass Spectrometry* (New York: Wiley-Interscience, 2001)

21. K W Busch, M A Busch (Eds) *Cavity-Ringdown Spectroscopy. An Ultratrace-Absorption Measurement Technique* (Washington, DC: American Chemical Society, 1999)

22. B Welz *Spectrochim. Acta, Part B* **54** 2081 (1999)

23. K W Jackson *Anal. Chem.* **72** R159 (2000)

24. M Schuetz, J Murphy, R E Fields, J M Harnly *Spectrochim. Acta, Part B* **55** 1895 (2000)

25. J M Harnly *J. Anal. At. Spectrom.* **14** 137 (1999)

26. B Welz, M G R Vale, M M Silva, H Becker-Ross, M D Huang, S Florek, U Heitmann *Spectrochim. Acta, Part B* **57** 1043 (2002)

27. A K Gilmutdinov, K Y Nagulin, M Sperling *J. Anal. At. Spectrom.* **15** 1375 (2000)

28. D J Butcher, A Zybin, M A Bolshov, K Niemax *Rev. Anal. Chem.* **20** 79 (2001)

29. A Zybin, J Koch, H D Wizemann, J Franzke, K Niemax *Spectrochim. Acta, Part B* **60** 1 (2005)

30. A A Bol'shakov, B A Cruden, S P Sharma *Proc. SPIE* **5339** 415 (2004)

31. P A Martin *Chem. Soc. Rev.* **31** 201 (2002)

32. P Werle, F Slemr, K Maurer, R Kormann, R Mücke, B Jänker *Opt. Lasers Eng.* **37** 101 (2002)

33. P Kluczynski, J Gustafsson, A M Lindberg, O Axner *Spectrochim. Acta, Part B* **56** 1277 (2001)

34. T Ban, H Skenderovic, S Ter-Avetisyan, G Pichler *Appl. Phys. B* **72** 337 (2001)

35. D Aumiler, T Ban, R Beuc, G Pichler *Appl. Phys.* B **76** 859 (2003)

36. V S Letokhov (Ed.) *Lazernaya Analiticheskaya Spektroskopiya* (Laser Analytical Spectroscopy) (Moscow: Nauka, 1986)

37. G Berden, R Peeters, G Meijer *Int. Rev. Phys. Chem.* **19** 565 (2000)

38. S S Brown *Chem. Rev.* **103** 5219 (2003)

39. D B Atkinson *Analyst (Cambridge, U.K.)* **128** 117 (2003)

40. J D Winefordner, I B Gornushkin, D Pappas, O I Matveev, B W Smith *J. Anal. At. Spectrom.* **15** 1161 (2000)

41. C Wang, F J Mazzotti, G P Miller, C B Winstead *Appl. Spectrosc.* **56** 386 (2002)

42. C J Wang, F J Mazzotti, G P Miller, C B Winstead *Appl. Spectrosc.* **57** 1167 (2003)

43. Y X Duan, C J Wang, C B Winstead *Anal. Chem.* **75** 2105 (2003)

44. A Schwabedissen, A Brockhaus, A Georg, J Engemann *J. Phys. D* **34** 1116 (2001)

45. W M M Kessels, J P M Hoefnagels, M G H Boogaarts, D C Schram, M C M van de Sanden *J. Appl. Phys.* **89** 2065 (2001)

46. V Bulatov, A Khalmanov, I Schechter *Anal. Bioanal. Chem.* **375** 1282 (2003)

47. A A Bol'shakov, N V Golovenkov, S V Oshemkov, A A Petrov *Spectrochim. Acta, Part B* **57** 355 (2002)

48. C B Winstead, F J Mazzotti, J Mierzwa, G P Miller *Anal. Commun.* **36** 277 (1999)

49. S Tao, F J Mazzotti, C B Winstead, G P Miller *Analyst (Cambridge, U.K.)* **125** 1021 (2000)

50. S Spuler, M Linne, A Sappey, S Snyder *Appl. Opt.* **39** 2480 (2000)

51. S Potapov, E Izrailov, V Vergizova, M Voronov, S Suprunovich, M Slyadnev, A Ganeev *J. Anal. At. Spectrom.* **18** 564 (2003)

52. D S Gough *Spectrochim. Acta, Part B* **54** 2067 (1999)

53. M Miclea, K Kunze, J Franzke, K Niemax *Spectrochim. Acta, Part B* **57** 1585 (2002)

54. A B Volynskii *Zh. Anal. Khim.* **58** 1015 (2003) [a]

55. A B Volynskii *Zh. Anal. Khim.* **59** 566 (2004) [a]

56. H M Ortner, E Bulska, U Rohr, G Schlemmer, S Weinbruch, B Welz *Spectrochim. Acta, Part B* **57** 1835 (2002)

57. Yu V Rogul'skii, R I Kholodov, L F Sukhodub *Zh. Anal. Khim.* **55** 360 (2000) [a]

58. R Oikari, V Haeyrinen, R Hernberg, L Kettunen *J. Anal. At. Spectrom.* **17** 1421 (2002)

59. H D Projahn, U Steeg, J Sanders, E Vanclay *Anal. Bioanal. Chem.* **378** 1083 (2004)

60. V V Bertsev, V B Borisov, V M Nemets, D S Skvortsov, A A Solov'ev *Zavod. Lab.* **68** 12 (2002)

61. L O Leal, N V Semenova, R Forteza, V Cerda *Talanta* **64** 1335 (2004)

62. N M M Coelho, A C da Silva, C M da Silva *Anal. Chim. Acta* **460** 227 (2002)

63. C Kabengera, P Bodart, P Hubert, L Thunus, A Noirfalise *J. AOAC Int.* **85** 122 (2002)

64. M Plessi, D Bertelli, A Monzani *J. Food Compos. Anal.* **14** 461 (2001)

65. B V L'vov *Spectrochim. Acta, Part B* **54** 2063 (1999)

66. B V L'vov *Spectrochim. Acta, Part B* **54** 1637 (1999)

67. A Bogaerts, E Neyts, R Gijbels, J van der Mullen *Spectrochim. Acta, Part B* **57** 609 (2002)

68. J A C Broekaert *Spectrochim. Acta, Part B* **55** 739 (2000)

69. M R Winchester, R Payling *Spectrochim. Acta, Part B* **59** 607 (2004)

70. J Angeli, A Bengtson, A Bogaerts, V Hoffmann, V-D Hodoroaba, E Steers *J. Anal. At. Spectrom.* **18** 670 (2003)

71. R K Marcus, E H Evans, J A Caruso *J. Anal. At. Spectrom.* **15** 1 (2000)

72. N P Zaksas, I R Shelpakova, V G Gerasimov *Zh. Anal. Khim.* **58** 254 (2003) [a]

73. A S Cherevko, G E Polyakova *Zh. Anal. Khim.* **60** 165 (2005) [a]

74. S Luan, R G Schleicher, M J Pilon, F D Bulman, G N Coleman *Spectrochim. Acta, Part B* **56** 1143 (2001)

75. H Becker-Ross, S Florek, H Franken, B Radziuk, M Zeiher *J. Anal. At. Spectrom.* **15** 851 (2000)

76. S Florek, C Haisch, M Okruss, H Becker-Ross *Spectrochim. Acta, Part B* **56** 1027 (2001)

77. J M Mermet *J. Anal. At. Spectrom.* **17** 1065 (2002)

78. B Rosenkranz, J Bettmer *Trends Anal. Chem. (Pers. Ed.)* **19** 138 (2000)

79. W J Yang, H Q Zhang, A M Yu, Q H Jin *Microchem. J.* **66** 147 (2000)

80. S Y Lu, C W LeBlanc, M W Blades *J. Anal. At. Spectrom.* **16** 256 (2001)

81. P Grinberg, R C Campos, Z Mester, R E Sturgeon *J. Anal. At. Spectrom.* **18** 902 (2003)

82. P Grinberg, R C Campos, R E Sturgeon *J. Anal. At. Spectrom.* **17** 693 (2002)

83. J Franzke, K Kunze, M Miclea, K Niemax *J. Anal. At. Spectrom.* **18** 802 (2003)

84. V Karanassios *Spectrochim. Acta, Part B* **59** 909 (2004)

85. J A C Broekaert *Anal. Bioanal. Chem.* **374** 182 (2002)

86. E Tognoni, V Palleschi, M Corsi, G Cristoforetti *Spectrochim. Acta, Part B* **57** 1115 (2002)

87. R E Russo, X Mao, H Liu, J Gonzalez, S S Mao *Talanta* **57** 423 (2002)

88. W B Lee, J Y Wu, Y I Lee, J Sneddon *Appl. Spectrosc. Rev.* **39** 27 (2004)

89. K Song, Y I Lee, J Sneddon *Appl. Spectrosc. Rev.* **37** 89 (2002)

90. D Günther, I Horn, B Hattendorf *Fresenius' J. Anal. Chem.* **368** 4 (2000)

91. R E Russo, X Mao, J J Gonzalez, S S Mao *J. Anal. At. Spectrom.* **17** 1072 (2002)

92. V Sturm, R Noll *Appl. Opt.* **42** 6221 (2003)

93. J D Hybl, G A Lithgow, S G Buckley *Appl. Spectrosc.* **57** 1207 (2003)

94. J E Carranza, B T Fisher, G D Yoder, D W Hahn *Spectrochim. Acta, Part B* **56** 851 (2001)

95. U Panne, R E Neuhauser, C Haisch, H Fink, R Niessner *Appl. Spectrosc.* **56** 375 (2002)

96. L Peter, V Sturm, R Noll *Appl. Opt.* **42** 6199 (2003)

97. R G Wiens, R E Arvidson, D A Cremers, M J Ferris, J D Blacic, F P Seelos, K S Deal *J. Geophys. Res., E (Planets)* **107** No. 8003 (2002)

98. A K Knight, N L Scherbarth, D A Cremers, M J Ferris *Appl. Spectrosc.* **54** 331 (2000)

99. J J Zayhowski, L A Wilson *IEEE J. Quantum Electron.* **38** 1449 (2002)

100. V I Tsarev, O A Bukin, S S Golik, A A Il'in, M V Trenina *Issledovano v Rossii* **6** 2108 (2003) (http://zhurnal.ape.relarn.ru/articles/2003/174.pdf)

101. S Palanco, A Alises, J Cunat, J Baena, J J Laserna *J. Anal. At. Spectrom.* **18** 933 (2003)

102. M D Cheng *Fuel Process. Technol.* **65** 219 (2000)

103. I B Gornushkin, K Amponsah-Manager, B W Smith, N Omenetto, J D Winefordner *Appl. Spectrosc.* **58** 762 (2004)

104. A A Antipov, A Z Grasyuk, S V Efimovskii, S V Kurbasov, L L Losev, V I Soskov *Kvant. Elektron.* **25** 31 (1998) [*Quant. Electr.* **28** 29 (1998)]

105. L St-Onge, V Detalle, M Sabsabi *Spectrochim. Acta, Part B* **57** 121 (2002)

106. S L Lui, N H Cheung *Appl. Phys. Lett.* **81** 5114 (2002)

107. L P Zhitenko, Yu V Taldonov, L E Ozerova, V P Obrezumov *Zavod. Lab.* **67** 22 (2001)

108. A J R Hunter, R T Wainner, L G Piper, S J Davis *Appl. Opt.* **42** 2102 (2003)

109. A J R Hunter, S J Davis, L G Piper, K W Holtzclaw, M E Fraser *Appl. Spectrosc.* **54** 575 (2000)

110. K Wagatsuma *Spectrochim. Acta, Part B* **56** 465 (2001)

111. K Shimizu, H Habazaki, P Skeldon, G E Thompson *Spectrochim. Acta, Part B* **58** 1573 (2003)

112. S Baude, J A C Broekaert, D Delfosse, N Jakubowski, L Fuechtjohann, N G Orellana-Velado, R Pereiro, A Sanz-Medel *J. Anal. At. Spectrom.* **15** 1516 (2000)

113. M Betti, L A de las Heras *Spectrochim. Acta, Part B* **59** 1359 (2004)

114. V S Burakov, V V Kiris, P A Naumenkov, S N Raikov *Zh. Prikl. Spektrosk.* **71** 676 (2004) [b]

115. D Bulajic, M Corsi, G Cristoforetti, S Legnaioli, V Palleschi, A Salvetti, E Tognoni *Spectrochim. Acta, Part B* **57** 339 (2002)

116. A A Bol'shakov, G G Glavin, R M Barnes *ICP Inf. Newslett.* **26** 136 (2000)

117. J S Becker, H-J Dietze *Int. J. Mass Spectrom. Ion Processes* **228** 127 (2003)

118. J S Becker, H-J Dietze *Int. J. Mass Spectrom. Ion Processes* **197** 1 (2000)

119. G M Hieftje, J H Barnes, O A Gron, A M Leach, D M McClenathan, S J Ray, D A Solyom, W C Wetzel, M B Denton, D W Koppenaal *Pure Appl. Chem.* **73** 1579 (2001)

120. J S Becker *Spectrochim. Acta, Part B* **57** 1805 (2002)

121. S N Ketkar, A D Scott, E J Hunter *Int. J. Mass Spectrom. Ion Processes* **206** 7 (2001)

122. M Kitano, Y Shirai, A Ohki, S Babasaki, T Ohmi *Jpn. J. Appl. Phys., Part 1* **40** 2688 (2001)

123. K P Jochum, B Stoll, J A Pfänder, M Seufert, M Flanz, P Maissenbacher, M Hofmann, A W Hofmann *Fresenius' J. Anal. Chem.* **370** 647 (2001)

124. N Nonose, M Kubota *J. Anal. At. Spectrom.* **16** 560 (2001)

125. M Moldovan, E M Krupp, A E Holliday, O F X Donard *J. Anal. At. Spectrom.* **19** 815 (2004)

126. Z Cheng, Y Zheng, R Mortlock, A van Geen *Anal. Bioanal. Chem.* **379** 512 (2004)

127. J M Marchante-Gayon *Anal. Bioanal. Chem.* **379** 335 (2004)

128. D R Bandura, V I Baranov, S D Tanner *Fresenius' J. Anal. Chem.* **370** 454 (2001)

129. D W Koppenaal, G C Eiden, C J Barinaga *J. Anal. At. Spectrom.* **19** 561 (2004)

130. R K Marcus *J. Anal. At. Spectrom.* **19** 591 (2004)

131. S D Tanner, V I Baranov, D Bandura *Spectrochim. Acta, Part B* **57** 1361 (2002)

132. D A Solyom, G M Hieftje *J. Am. Soc. Mass Spectrom.* **14** 227 (2003)

133. D A Solyom, T W Burgoyne, G M Hieftje *J. Anal. At. Spectrom.* **14** 1101 (1999)

134. E Engström, A Stenberg, D C Baxter, D Malinovsky, I Mäkinen, S Pönni, I Rodushkin *J. Anal. At. Spectrom.* **19** 858 (2004)

135. S J Ray, G M Hieftje *J. Anal. At. Spectrom.* **16** 1206 (2001)

136. M Balcerzak *Anal. Sci.* **19** 979 (2003)

137. C S Westphal, J A McLean, B W Acon, L A Allen, A Montaser *J. Anal. At. Spectrom.* **17** 669 (2002)

138. D M McClenathan, S J Ray, W C Wetzel, G M Hieftje *Anal. Chem.* **76** 158A (2004)

139. B Hattendorf, C Latkoczy, D Gunther *Anal. Chem.* **75** 341A (2003)

140. J H Barnes, G D Schilling, G M Hieftje, R P Sperline, M B Denton, C J Barinaga, D W Koppenaal *J. Am. Soc. Mass Spectrom.* **15** 769 (2004)

141. A M Leach, G M Hieftje *J. Anal. At. Spectrom.* **17** 852 (2002)

142. V S Sevast'yanov, G I Ramendik, S M Sil'nov, N E Babulevich, D A Tyurin, E V Fatyushina *Zh. Anal. Khim.* **57** 509 (2002) [a]

143. A A Sysoev, S S Poteshin, A Yu Adamov, I V Martynov *Zavod. Lab.* **67** 16 (2001)

144. A A Sysoev, A A Sysoev *Eur. J. Mass Spectrom.* **8** 213 (2002)

145. A A Sysoev, S S Poteshin, G B Kuznetsov, I A Kovalev, E S Yushkov *Zh. Anal. Khim.* **57** 958 (2002) [a]

146. V V Bezrukov, I D Kovalev, K N Malyshev, D K Ovchinnikov *Zh. Anal. Khim.* **59** 723 (2004) [a]

147. U Rohner, J A Whitby, P Wurz *Meas. Sci. Technol.* **14** 2159 (2003)

148. M Yorozu, T Yanagida, T Nakajyo, Y Okada, A Endo *Appl. Opt.* **40** 2043 (2001)

149. K Watanabe, K Hattori, J Kawarabayashi, T Iguchi *Spectrochim. Acta, Part B* **58** 1163 (2003)

150. D M Grim, J Allison *Int. J. Mass Spectrom. Ion Processes* **222** 85 (2003)

151. M V Peláez, J M Costa-Fernández, R Pereiro, N Bordel, A Sanz-Medel *J. Anal. At. Spectrom.* **18** 864 (2003)

152. J Pisonero, J M Costa, R Pereiro, N Bordel, A Sanz-Medel *Anal. Bioanal. Chem.* **379** 658 (2004)

153. A A Ganeev, M A Kuzmenkov, M N Slyadnev, M V Voronov, S V Potapov, in *The Pittsburgh Conference on Analytical Chemistry and Applied Spectroscopy — Pittcon 2005 (Abstracts of Reports), Orlando, 2005* p. 53

154. A Chatterjee, Y Shibata, A Tanaka, M Morita *Anal. Chim. Acta* **436** 253 (2001)

155. A Chatterjee, Y Shibata, J Yoshinaga, A Tanaka, M Morita *Anal. Chem.* **72** 4402 (2000)

156. I J Stewart, R E Sturgeon *J. Anal. At. Spectrom.* **15** 1223 (2000)

157. K H Rubin *Chem. Geol.* **175** 723 (2001)

158. L Pajo, G Tamborini, G Rasmussen, K Mayer, L Koch *Spectrochim. Acta, Part B* **56** 541 (2001)

159. P Ghosh, W A Brand *Int. J. Mass Spectrom. Ion Processes* **228** 1 (2003)

160. L R Karam, L Pibida, C A McMahon *Appl. Radiat. Isot.* **56** 369 (2002)

161. S S Dimov, S L Chryssoulis, R H Lipson *Anal. Chem.* **75** 6723 (2003)

162. M R Savina, M J Pellin, C E Tripa, I V Veryovkin, W F Calaway, A M Davis *Geochim. Cosmochim. Acta* **17** 3215 (2003)

163. K Blaum, C Geppert, W G Schreiber, J G Hengstler, P Müller, W Nörtershäuser, K Wendt, B A Bushaw *Anal. Bioanal. Chem.* **372** 759 (2002)

164. M J Pellin, W F Calaway, A M Davis, R S Lewis, S Amari, R N Clayton, in *The 31st Lunar and Planetary Science Conference (Abstracts of Reports), Houston, TX, 2000* p. 1917

165. U Köster, V N Fedoseyev, V I Mishin *Spectrochim. Acta, Part B* **58** 1047 (2003)

166. K D A Wendt *Eur. J. Mass Spectrom.* **8** 273 (2002)

167. N Trautmann, G Passler, K D A Wendt *Anal. Bioanal. Chem.* **378** 348 (2004)

168. M Suter *Nucl. Instrum. Methods Phys. Res., Sect. B* **223/224** 139 (2004)

169. R Hellborg, M Faarinen, M Kiisk, C E Magnusson, P Persson, G Skog, K Stenström *Vacuum* **70** 365 (2003)

170. L K Fifield *Nucl. Instrum. Methods Phys. Res., Sect. B* **223/224** 401 (2004)

171. P Muzikar, D Elmore, D E Granger *Geol. Soc. Am. Bull.* **115** 643 (2003)

172. C Tuniz, U Zoppi, M A C Hotchkis *Nucl. Instrum. Methods Phys. Res., Sect. B* **213** 469 (2004)

173. F Li, Z Xie, H Schmidt, S Sielemann, J I Baumbach *Spectrochim. Acta, Part B* **57** 1563 (2002)

174. R G Ewing, D A Atkinson, G A Eiceman, G J Ewing *Talanta* **54** 515 (2001)

175. S N Ketkar, S Dheandhanoo *Anal. Chem.* **73** 2554 (2001)

176. R G Longwitz, H van Lintel, P Renaud *J. Vac. Sci. Technol. B* **21** 1570 (2003)

177. I A Buryakov, Yu N Kolomiets, V B Luppu *Zh. Anal. Khim.* **56** 381 (2001)[a]

178. R Clough, J Truscatt, S T Belt, E H Evans, B Fairman, T Catterick *Appl. Spectrosc. Rev.* **38** 101 (2003)

179. K G Heumann *Anal. Bioanal. Chem.* **378** 318 (2004)

180. A A Pupyshev, A K Lutsak *Anal. Kontrol'* **4** 304 (2000)

181. G I Ramendik, E V Fatyushina, A I Stepanov, V S Sevast'yanov *Zh. Anal. Khim.* **56** 566 (2001)[a]

182. G I Ramendik, E V Fatyushina, A I Stepanov *Zh. Anal. Khim.* **57** 20 (2002)[a]

183. G I Ramendik, E V Fatyushina, A I Stepanov *Zh. Anal. Khim.* **58** 174 (2003)[a]

184. N A Klyuev *Zh. Anal. Khim.* **57** 566 (2002)[a]

185. V A Khvostikov, G F Telegin, S S Grazhulene *Zh. Anal. Khim.* **58** 581 (2003)[a]

186. H Bagheri, A Gholami *Talanta* **55** 1141 (2001)

187. L Liang, M Horvat, H Li, P Pang *J. Anal. At. Spectrom.* **18** 1383 (2003)

188. A D'Ulivo, E Bramanti, L Lampugnani, R Zamboni *Spectrochim. Acta, Part B* **56** 1893 (2001)

189. N V Semenova, L O Leal, R Forteza, V Cerda *Anal. Chim. Acta* **486** 217 (2003)

190. A Sayago, R Beltrán, M A F Recamales, J L Gómez-Ariza *J. Anal. At. Spectrom.* **17** 1440 (2002)

191. S Jianbo, T Zhiyong, T Chunchua, C Quan, J Zexiang *Talanta* **56** 711 (2002)

192. H Sun, R Suo, Y Lu *Anal. Chim. Acta* **457** 305 (2002)

193. J T van Elteren, Z Slejkovec, H A Das *Spectrochim. Acta, Part B* **54** 311 (1999)

194. Aurora AI-3200: www.aurora-instr.com/AI3200NF.htm

195. A Young, L Pitts, S Greenfield, M Foulkes *J. Anal. At. Spectrom.* **18** 44 (2003)

196. P Stchur, K X Yang, X Hou, T Sun, R G Michel *Spectrochim. Acta, Part B* **56** 1565 (2001)

197. Y He, B J Orr *Appl. Opt.* **40** 4836 (2001)

198. J Mes, M Leblans, W Hogervorst *Opt. Lett.* **27** 1442 (2002)

199. G W Baxter, M A Payne, B D W Austin, C A Halloway, J G Haub, Y He, A P Milce, J F Nibler, B J Orr *Appl. Phys. B* **71** 651 (2000)

200. N P Barnes, in *Tunable Lasers Handbook* (Ed. F J Duarte) (San Diego, CA: Academic Press, 1995) p. 293

201. L Burel, P Giamarchi, L Stephan, Y Lijour, A Le Bihan *Talanta* **60** 295 (2003)

202. J X Zhou, X Hou, K X Kang, R G Michel *J. Anal. At. Spectrom.* **13** 41 (1998)

203. S Aubin, E Gomez, L A Orozco, G D Sprouse *Rev. Sci. Instrum.* **74** 4342 (2003)

204. K Bailey, C Y Chen, X Du, Y M Li, Z T Lu, T P O'Connor, L Young *Nucl. Instrum. Methods Phys. Res., Sect. B* **172** 224 (2000)

205. I D Moore, K Bailey, J Greene, Z T Lu, P Müller, T P O'Connor, C Geppert, K D A Wendt, L Young *Phys. Rev. Lett.* **92** 153002/1 (2004)

206. O M Marago, B Fazio, P G Gucciardi, E Arimondo *Appl. Phys. B* **77** 809 (2003)

207. P E Walters, T E Barber, M W Wensing, J D Winefordner *Spectrochim. Acta, Part B* **46** 1015 (1991)

208. A Zybin, C Schnürer-Patschan, K Niemax *Spectrochim. Acta, Part B* **47** 1519 (1992)

209. K Yuasa, Y Yamashina, T Sakurai *Jpn. J. Appl. Phys.* **36** 2340 (1997)

210. C Raab, J Bolle, H Oberst, J Eschner, F Schmidt-Kaler, R Blatt *Appl. Phys. B* **67** 683 (1998)

211. G D Severn, D A Edrich, R McWilliams *Rev. Sci. Instrum.* **69** 10 (1998)

212. B W Smith, A Quentmeier, M Bolshov, K Niemax *Spectrochim. Acta, Part B* **54** 943 (1999)

213. J P Temirov, O I Matveev, B W Smith, J D Winefordner *Appl. Spectrosc.* **57** 729 (2003)

214. W L Clevenger, O I Matveev, S Cabredo, N Omenetto, B W Smith, J D Winefordner *Anal. Chem.* **69** 2232 (1997)

215. W L Clevenger, L S Mordoh, O I Matveev, N Omenetto, B W Smith, J D Winefordner *Spectrochim. Acta, Part B* **52** 295 (1997)

216. J F Gravel, P Nobert, J F Y Gravel, D Boudreau *Anal. Chem.* **75** 1442 (2003)

217. H L Pacquette, S A Elwood, M Ezer, J B Simeonsson *J. Anal. At. Spectrom.* **16** 152 (2001)

218. K L Riter, O I Matveev, B W Smith, J D Winefordner *Anal. Chim. Acta* **333** 187 (1996)

219. C-B Ke, K-C Lin *Anal. Chem.* **71** 1561 (1999)

220. C-B Ke, K-D Su, K-C Lin *J. Chromatogr. A* **921** 247 (2001)

221. A A Gorbatenko, N B Zorov, T A Labutin *Zh. Anal. Khim.* **58** 388 (2003)[a]

222. A A Gorbatenko, N B Zorov, A R Murtazin *Zh. Anal. Khim.* **57** 151 (2002)[a]

223. J F Y Gravel, M L Viger, P Nobert, D Boudreau *Appl. Spectrosc.* **58** 727 (2004)

224. A T Khalmanov, D-K Ko, J Lee, N Eshkobilov, A Tursunov *J. Korean Phys. Soc.* **44** 843 (2004)

225. J B Simeonsson, S A Elwood, M Ezer, H L Pacquette, D J Swart, H D Beach, D J Thomas *Talanta* **58** 189 (2002)
226. A T Khalmanov, Kh S Khamraev, A T Tursunov, O Tukhliboev *Opt. Spektrosk.* **92** 885 (2002)[c]
227. A T Khalmanov, Kh S Khamraev, A T Tursunov, O Tukhliboev *Opt. Spektrosk.* **90** 400 (2001)[c]
228. D Boudreau, J F Gravel *Trends Anal. Chem. (Pers. Ed.)* **20** 20 (2001)
229. J C Travis, G C Turk (Eds) *Laser-Enhanced Ionization Spectrometry* (New York: Wiley, 1996)
230. B Barbieri, N Beverini, A Sasso *Rev. Mod. Phys.* **62** 603 (1990)
231. X Yao, S P McGlynn, R C Mohanty *Microchem. J.* **61** 223 (1999)
232. A T Khalmanov *Zh. Prikl. Spektrosk.* **67** 396 (2000)[b]
233. J P Young, R W Shaw, C M Barshick, J M Ramsey *J. Alloys Compd.* **271–273** 62 (1998)
234. E C Jung, T-S Kim, K Song, C-J Kim *Spectrosc. Lett.* **36** 167 (2003)
235. T J McCarthy, M Salvermoser, D E Murnick *Hyperfine Interact.* **151/152** 209 (2003)
236. A A Ganeev, V N Grigor'yan, A I Drobyshev, M N Slyadnev, S E Sholupov *Zh. Anal. Khim.* **51** 848 (1996)[a]
237. D L Monts, Abhilasha, S Qian, D Kumar, X Yao, S P McGlynn *J. Thermophys. Heat Transfer* **12** 66 (1998)
238. E de Oliveira *J. Braz. Chem. Soc.* **14** 174 (2003)
239. J Pawliszyn *Anal. Chem.* **75** 2543 (2003)
240. I V Kubrakova *Usp. Khim.* **71** 327 (2002) [*Russ. Chem. Rev.* **71** 283 (2002)]
241. J A Nóbrega, L C Trevizan, G C L Araújo, A R A Nogueira *Spectrochim. Acta, Part B* **57** 1855 (2002)
242. Yu A Zolotov, G I Tsizin, E I Morosanova, S G Dmitrienko *Usp. Khim.* **74** 41 (2005) [*Russ. Chem. Rev.* **74** 37 (2005)]
243. J C A Wuilloud, R G Wuilloud, A P Vonderheide, J A Caruso *Spectrochim. Acta, Part B* **59** 755 (2004)
244. M D Pereira, M A Z Arruda *Mikrochim. Acta* **141** 115 (2003)
245. K G Heumann *Anal. Bioanal. Chem.* **373** 323 (2002)
246. A I Saprykin, I R Shelpakova, T A Chanysheva, N P Zaksas *Zh. Anal. Khim.* **58** 273 (2003)[a]
247. M A Bolshov, A V Zybin, D Butcher, K Niemaks *Zh. Anal. Khim.* **58** 725 (2003)[a]
248. X-P Yan, Y Jiang *Trends Anal. Chem. (Pers. Ed.)* **20** 552 (2001)
249. J H Wang, E H Hansen *Trends Anal. Chem. (Pers. Ed.)* **22** 225 (2003)
250. R C Prados-Rosales, J L Luque-Garcia, M D L de Castro *Anal. Chim. Acta* **480** 181 (2003)
251. J H Wang, E H Hansen *Trends Anal. Chem. (Pers. Ed.)* **22** 836 (2003)
252. A D'Ulivo, C Baiocchi, E Pitzalis, M Onor, R Zamboni *Spectrochim. Acta, Part B* **59** 471 (2004)
253. Y-L Feng, R E Sturgeon, J W Lam *J. Anal. At. Spectrom.* **18** 1435 (2003)
254. E Bolea, F Laborda, J R Castillo, R E Sturgeon *Spectrochim. Acta, Part B* **59** 505 (2004)
255. P Smichowski, S V Farias *Microchem. J.* **67** 147 (2000)
256. X C Duan, R L McLaughlin, I D Brindle, A Conn *J. Anal. At. Spectrom.* **17** 227 (2002)
257. M Resano, M Verstraete, F Vanhaecke, L Moens *J. Anal. At. Spectrom.* **16** 1018 (2001)
258. M J Cal-Prieto, M Felipe-Sotelo, A Carlosena, J M Andrade, P Lopez-Mahia, S Muniategui, D Prada *Talanta* **56** 1 (2002)
259. T Kántor *Spectrochim. Acta, Part B* **56** 1523 (2001)
260. A S Luna, H B Pereira, I Takase, R A Goncalves, R E Sturgeon, R C de Campos *Spectrochim. Acta, Part B* **57** 2047 (2002)
261. J L Burguera, M Burguera *Spectrochim. Acta, Part B* **56** 1801 (2001)
262. X Hou, B T Jones *Spectrochim. Acta, Part B* **57** 659 (2002)
263. J A Nóbrega, J Rust, C P Calloway, B T Jones *Spectrochim. Acta, Part B* **59** 1337 (2004)
264. M Huang, H Kojima, T Shirasaki, A Hirabayashi, H Koizumi *Anal. Chim. Acta* **413** 217 (2000)
265. H Matusiewicz *Spectrochim. Acta, Part B* **57** 485 (2002)
266. A L Rosen, G M Hieftje *Spectrochim. Acta, Part B* **59** 135 (2004)

267. P J Dyson, A K Hearley, B F G Johnson, J S McIndoe, P R R Langridge-Smith, C Whyte *Rapid Commun. Mass Spectrom.* **15** 895 (2001)
268. I I Stewart *Spectrochim. Acta, Part B* **54** 1649 (1999)
269. P Rychlovský, P Černoch, M Sklenicková *Anal. Bioanal. Chem.* **374** 955 (2002)
270. B Lavine, J J Workman *Anal. Chem.* **76** 3365 (2004)
271. P Geladi *Spectrochim. Acta, Part B* **58** 767 (2003)
272. P Geladi, B Sethson, J Nyström, T Lillhonga, T Lestander, J Burger *Spectrochim. Acta, Part B* **59** 1347 (2004)
273. C M Wehlburg, D M Haaland, D K Melgaard *Appl. Spectrosc.* **56** 877 (2002)
274. H S Yang, P R Griffiths, J D Tate *Anal. Chim. Acta* **489** 125 (2003)
275. M A Proskurnin, M Yu Kononets *Usp. Khim.* **73** 1235 (2004) [*Russ. Chem. Rev.* **73** 1143 (2004)]
276. R E Russo, X L Mao, C Liu, J Gonzalez *J. Anal. At. Spectrom.* **19** 1084 (2004)
277. J L Todoli, J M Mermet *Trends Anal. Chem. (Pers. Ed.)* **24** 107 (2005)
278. J A C Broekaert, V Siemens *Spectrochim. Acta, Part B* **59** 1823 (2004)
279. V Hoffmann, M Kasik, P K Robinson, C Venzago *Anal. Bioanal. Chem.* **381** 173 (2005)

[a] — *J. Anal. Chem. (Engl. Transl.)*
[b] — *J. Appl. Spectrosc. (Engl. Transl.)*
[c] — *Opt. Spectrosc. (Engl. Transl.)*

**Same reference list with titles included:**

1. E.H. Evans, J.A. Day, A. Fisher, W.J. Price, C.M.M. Smith, J.F. Tyson, Atomic spectrometry update. Advances in atomic emission, absorption and fluorescence spectrometry and related techniques, *J. Anal. At. Spectrom.*, 2004, v. 19, p. 775-812.
2. E.H. Evans, J.A. Day, W.J. Price, C.M.M. Smith, K. Sutton, J.F. Tyson, Atomic spectrometry update. Advances in atomic emission, absorption and fluorescence spectrometry and related techniques, *J. Anal. At. Spectrom.*, 2003, v. 18, p. 808-833.
3. J.R. Bacon, J.C. Greenwood, L. van Vaeck, J.G. Williams, Atomic spectrometry update. Atomic mass spectrometry, *J. Anal. At. Spectrom.*, 2004, v. 19, p. 1020-1057.
4. S.J. Hill, T.A. Arowolo, O.T. Butler, J.M. Cook, M.S. Cresser, C. Harrington, D.L. Miles, Atomic spectrometry update. Environmental analysis, *J. Anal. At. Spectrom.*, 2004, v. 19, p. 301-330.
5. A. Fisher, P.S. Goodall, M.W. Hinds, S.M. Nelms, D.M. Penny, Atomic spectrometry update. Industrial analysis: metals, chemicals and advanced materials, *J. Anal. At. Spectrom.*, 2004, v. 19, p. 1567-1595.
6. A. Taylor, S. Branch, D. Halls, M. Patriarca, M. White, Atomic spectrometry update. Clinical and biological materials, foods and beverages, *J. Anal. At. Spectrom.*, 2004, v. 19, p. 505-556.
7. N.H. Bings, A. Bogaerts, J.A.C. Broekaert, Atomic spectroscopy, *Anal. Chem.*, 2004, v. 76, p. 3313-3336.
8. N.H. Bings, A. Bogaerts, J.A.C. Broekaert, Atomic spectroscopy, *Anal. Chem.*, 2002, v. 74, p. 2691-2711.
9. J.D. Winefordner, I.B. Gornushkin, T. Correll, E. Gibb, B.W. Smith, N. Omenetto, Comparing several atomic spectrometric methods to the super stars: special emphasis on laser induced breakdown spectrometry, LIBS, a future super star, *J. Anal. At. Spectrom.*, 2004, v. 19, p. 1061-1083.
10. А.А. Пупышев, А.К. Луцак, Современное состояние методов атомного спектрального анализа, *Аналит. контроль*, 2000, т. 4, c. 141-146.
11. М.А. Большов, Некоторые современные методы инструментального элементного анализа и тенденции их развития, *Завод. лабор.*, 2004, т. 70, c. 3-13.
12. M. Cullen, Editor, *Atomic Spectroscopy in Elemental Analysis*, Blackwell Publishing, Oxford (UK); CRC Press, Boca Raton (USA, FL) 2004, 328 p.
13. J. Sneddon, Editor, *Advances in Atomic Spectroscopy*, v. 7, Elsevier Science, Amsterdam, 2002, 406 p.
14. А.А. Пупышев, Д.А. Данилова, *Атомно-эмиссионный спектральный анализ с индуктивно связанной плазмой и тлеющим разрядом по грримму*, ГОУ ВПО УГТУ-УПИ, Екатеринбург, 2002, 202 c.
15. B. Welz, M. Sperling, *Atomic Absorption Spectrometry*, Wiley-VCH, Weinheim (Germany), 3rd ed., 1999, 941 p.
16. J.A.C. Broekaert, *Analytical Atomic Spectrometry with Flames and Plasmas*, Wiley-VCH, Weinheim (Germany), 2002, 376 p.
17. R.K. Marcus, J.A.C. Broekaert, Editors, *Glow Discharge Plasmas in Analytical Spectroscopy*, J. Wiley & Sons, Chichester (UK), 2003, 498 p.
18. Y.I. Lee, K. Song, J. Sneddon, *Laser-Induced Breakdown Spectrometry*, Nova Science Publishers, Huntington (USA, NY), 2000, 180 p.
19. P. Silvester, Editor, *Laser Ablation-ICP-MS in the Earth Sciences*, Mineralogical Association of Canada, St. John's (Newfoundland), 2001, 243 p.
20. J.R. de Laeter, *Applications of Inorganic Mass Spectrometry*, J. Wiley (Wiley Interscience Ser.), New York, 2001, 474 p.
21. K.W. Busch, M.A. Busch, Editors, *Cavity-Ringdown Spectroscopy. An Ultratrace-Absorption Measurement Technique*, American Chemical Society, Washington (USA, DC), 1999, 270 p.
22. B. Welz, Atomic absorption spectroscopy – pregnant again after 45 years, *Spectrochim. Acta, Part B*, 1999, v. 54, p. 2081-2094.
23. K.W. Jackson, Electrothermal atomic absorption spectrometry and related techniques, *Anal. Chem.*, 2000, v. 72, p. R159-R167.
24. M. Schuetz, J. Murphy, R.E. Fields, J.M. Harnly, Continuum source-atomic absorption spectrometry using a two-dimensional charge coupled device, *Spectrochim. Acta, Part B*, 2000, v. 55, p. 1895-1912.
25. J.M. Harnly, The future of atomic absorption spectrometry: a continuum source with a charge coupled array detector, *J. Anal. At. Spectrom.*, 1999, v. 14, p. 137-146.
26. B. Welz, M.G.R. Vale, M.M. Silva, H. Becker-Ross, M.D. Huang, S. Florek, U. Heitmann, Investigation of interferences in the determination of thallium in marine sediment reference materials using high-resolution continuum-source atomic absorption spectrometry and electrothermal atomization, *Spectrochim. Acta, Part B*, 2002, v. 57, p. 1043-1055.
27. A.K. Gilmutdinov, K.Y. Nagulin, M. Sperling, Spatially resolved atomic absorption analysis, *J. Anal. At. Spectrom.*, 2000, v. 15, p. 1375-1382.
28. D.J. Butcher, A. Zybin, M.A. Bolshov, K. Niemax, Diode laser atomic absorption spectrometry as a detector for metal speciation, *Rev. Anal. Chem.*, 2001, v. 20, p. 79-100.
29. A. Zybin, J. Koch, H.D. Wizemann, J. Franzke, K. Niemax, Diode laser atomic absorption spectrometry, *Spectrochim. Acta, Part B*, 2005, v. 60, p. 1-11.
30. A.A. Bol'shakov, B.A. Couden, S.P. Sharma, Sensor for monitoring plasma parameters, *Proc. SPIE*, 2004, v. 5339, p. 415-426.
31. P.A. Martin, Near-infrared diode laser spectroscopy in chemical process and environmental air monitoring, *Chem. Soc. Rev.*, 2002, v. 31, p. 201-210.
32. P. Werle, F. Slemr, K. Maurer, R. Kormann, R. Mücke, B. Jänker, Near- and mid-infrared laser-optical sensors for gas analysis, *Opt. Lasers Eng.*, 2002, v. 37, p. 101-114.
33. P. Kluczynski, J. Gustafsson, A.M. Lindberg, O. Axner, Wavelength modulation absorption spectrometry - an extensive scrutiny of the generation of signals, *Spectrochim. Acta, Part B*, 2001, v. 56, p. 1277-1354.
34. T. Ban, H. Skenderovic, S. Ter-Avetisyan, G. Pichler, Absorption measurements in dense cesium vapor using a UV-violet light-emitting diode, *Appl. Phys. B*, 2001, v. 72, p. 337-341.
35. D. Aumiler, T. Ban, R. Beuc, G. Pichler, Simultaneous determination of the temperature and density of rubidium vapor, *Appl. Phys. B*, 2003, v. 76, p. 859-867.
36. Лазерная аналитическая спектроскопия. Под ред. В.С. Летохова. Наука, Москва, 1986, 316 c.
37. G. Berden, R. Peeters, G. Meijer, Cavity ring-down spectroscopy: Experimental schemes and applications, *Int. Rev. Phys. Chem.*, 2000, v. 19, p. 565-607.
38. S.S. Braun, Absorption spectroscopy in high-finesse cavities for atmospheric studies, *Chem. Rev.*, 2003, v. 103, p. 5219-5238.
39. D.B. Atkinson, Solving chemical problems of environmental importance using cavity ring-down spectroscopy, *Analyst*, 2003, v. 128, p. 117-125.
40. J.D. Winefordner, I.B. Gornushkin, D. Pappas, O.I. Matveev, B.W. Smith, Novel uses of lasers in atomic spectroscopy, *J. Anal. At. Spectrom.*, 2000, v. 15, p. 1161-1189.
41. C. Wang, F.J. Mazzotti, G.P. Miller, C.B. Winstead, Cavity ringdown spectroscopy for diagnostic and analytical measurements in an inductively coupled plasma, *Appl. Spectrosc.*, 2002, v. 56, p. 386-397.
42. C.J. Wang, F.J. Mazzotti, G.P. Miller, C.B. Winstead, Isotopic measurements of uranium using inductively coupled plasma cavity ringdown spectroscopy, *Appl. Spectrosc.*, 2003, v. 57, p. 1167-1172.
43. Y.X. Duan, C.J. Wang, C.B. Winstead, Exploration of microwave plasma source cavity ring-down spectroscopy for elemental measurements, *Anal. Chem.*, 2003, v. 75, p. 2105-2111.
44. A. Schwabedissen, A. Brockhaus, A. Georg, J. Engemann, Determination of the gas-phase Si atom density in radio frequency discharges by means of cavity ring-down spectroscopy, *J. Phys. D.*, 2001, v. 34, p. 1116-1121.
45. W.M.M. Kessels, J.P.M. Hoefnagels, M.G.H. Boogaarts, D.C. Schram, M.C.M. van de Sanden, Cavity ring down study of the densities and kinetics of Si and SiH in a remote Ar-H$_2$-SiH$_4$ plasma, *J. Appl. Phys.*, 2001, v. 89, p. 2065-2073.

46. V. Bulatov, A. Khalmanov, I. Schechter, Study of the morphology of laser-produced aerosol plume by cavity ringdown laser absorption spectroscopy, *Anal. Bioanal. Chem.*, 2003, v. 375, p. 1282-1286.

47. A.A. Bol'shakov, N.V. Golovenkov, S.V. Oshemkov, A.A. Petrov, Corrective correlation method for the analysis of gases by plasma/laser–excited fluorescence spectrometry, *Spectrochim. Acta, Part B*, 2002, v. 57, p. 355-364.

48. C.B. Winstead, F.J. Mazzotti, J. Mierzwa, G.P. Miller, Preliminary results for electrothermal atomization-cavity ringdown spectroscopy (ETA-CRDS), *Anal. Commun.*, 1999, v. 36, p. 277-279.

49. S. Tao, F.J. Mazzotti, C.B. Winstead, G.P. Miller, Determination of elemental mercury by cavity ringdown spectrometry, *Analyst*, 2000, v. 125, p. 1021-1023.

50. S. Sputter, M. Linne, A. Sappey, S. Snyder, Development of a cavity ringdown laser absorption spectrometer for detection of trace level of mercury, *Appl. Opt.*, 2000, v. 39, p. 2480-2486.

51. S. Potapov, E. Izrailov, V. Vergizova, M. Voronov, S. Suprunovich, M. Slyadnev, A. Ganeev, Pulsed glow discharge in thin-walled metallic hollow cathode. Analytical possibilities in atomic and mass spectrometry, *J. Anal. At. Spectrom.,* 2003, v. 18, p. 564-571.

52. D.S. Gough, The development of sputtering techniques and their application to atomic absorption spectrometry, *Spectrochim. Acta, Part B*, v. 54, 1999, p. 2067-2072.

53. M. Miclea, K. Kunze, J. Franzke, K. Niemax, Plasma for lab-on-the-chip applications, *Spectrochim. Acta, Part B*, 2002, v. 57, p. 1585-1592.

54. А.Б. Волынский, Химические модификаторы в современной электротермической атомно-абсорбционной спектрометрии, *Ж. аналит. хим.*, 2003, т. 58, с. 1015-1033.

55. А.Б. Волынский, Химические модификаторы на основе соединений платиновых металлов в электротермической атомно-абсорбционной спектрометрии, *Ж. аналит. хим.*, 2004, т. 59, с. 566-586.

56. H.M. Ortner, E. Bulska, U. Rohr, G. Schlemmer, S. Weinbruch, B. Welz, Modifiers and coatings in graphite furnace atomic absorption spectrometry – mechanisms of action (A tutorial review), *Spectrochim. Acta, Part B*, 2002, v. 57, p. 1835-1853.

57. Ю.В. Рогульский, Р.И. Холодов, Л.Ф. Суходуб, Кинетическая модель атомно-абсорбционного сигнала, *Ж. аналит. хим.*, 2000, т. 55, с. 360-366.

58. R. Oikari, V. Haeyrinen, R. Hernberg, L. Kettunen, Application of Zeeman background correction to direct current plasma excited atomic absorption spectroscopy, *J. Anal. At. Spectrom.*, 2002, v. 17, p. 1421-1424.

59. H.D. Projahn, U. Steeg, J. Sanders, E. Vanclay, Application of the reference-element technique for fast sequential flame atomic-absorption spectrometry, *Anal. Bioanal. Chem.*, 2004, v. 378, p. 1083-1087.

60. В.В. Берцев, В.Б. Борисов, В.М. Немец, Д.С. Скворцов, А.А. Соловьёв, О возможности применения метода главных компонент в аналитической абсорбционной спектроскопии, *Завод. лабор.*, 2002, т. 68, с. 12-16.

61. L.O. Leal, N.V. Semenova, R. Forteza, V. Cerdà, Preconcentration and determination of inorganic arsenic using a multisyringe flow injection system and hydride generation-atomic fluorescence spectrometry, *Talanta*, 2004, v. 64, p. 1335-1342.

62. N.M.M. Coelho, A.C. da Silva, C.M. da Silva, Determination of As (III) and total inorganic arsenic by flow injection hydride generation atomic absorption spectrometry, *Anal. Chim. Acta*, 2002, v. 460, p. 227-233.

63. C. Kabengera, P. Bodart, P. Hubert, L. Thunus, A. Noirfalise, Optimization and validation of arsenic determination in foods by hydride generation flame atomic absorption spectrometry, *J. AOAC Int.*, 2002, v. 85, p. 122-127.

64. M. Plessi, D. Bertelli, A. Monzani, Mercury and selenium content in selected seafood, *J. Food Compos. Anal.*, 2001, v. 14, p. 461-467.

65. B.V. L'vov, Alan Walsh and the absolute analysis project, *Spectrochim. Acta, Part B*, 1999, v. 54, p. 2063-2065.

66. B.V. L'vov, A continuum source vs. line source on the way toward absolute graphite furnace atomic absorption spectrometry, *Spectrochim. Acta, Part B*, 1999, v. 54, p. 1637-1646.

67. A. Bogaerts, E. Neyts, R. Gijbelts, J. van der Mullen, Gas discharge plasmas and their applications, *Spectrochimica Acta, Part B*, v. 57, 2002, p. 609-658.

68. J.A.C. Broekaert, State-of-the-art and trends of development in analytical atomic spectrometry with inductively coupled plasmas as radiation and ion sources, *Spectrochim. Acta, Part B*, 2000, v. 55, p. 739-751.

69. M.R. Winchester, R. Payling, Radio-frequency glow discharge spectrometry: A critical review, *Spectrochim. Acta, Part B*, 2004, v. 59, p. 607-666.

70. J. Angeli, A. Bengtson, A. Bogaerts, V. Hoffmann, V.-D. Hodoroaba, E. Steers, Glow discharge optical emission spectrometry: moving towards reliable thin film analysis – a short review, *J. Anal. At. Spectrom.*, 2003, v. 18, p. 670-679.

71. R.K. Marcus, E.H. Evans, J.A.Caruso, Tunable plasma sources in analytical spectroscopy: Current status and projections, *J. Anal. At. Spectrom.*, 2000, v. 15, p. 1-5.

72. Н.П. Заксас, И.Р. Шелпакова, В.Г. Герасимов, Атомно-эмиссионное определение микроэлементов в порошковых пробах разной природы с возбуждением спектров в двухструйном дуговом плазматроне, *Ж. аналит. хим.*, 2003, т. 58, с. 254-261.

73. А.С. Черевко, Г.Е. Полякова, Атомно-эмиссионное спектрографическое определение валового содержания таллия в почвах с дуговым аргоновым двухструнным плазматроном, *Ж. аналит. хим.*, 2005, т. 60, с. 165.

74. S. Luan, R.G. Schleicher, M.J. Pilon, F.D. Bulman, G.N. Coleman, An echelle polychromator for inductively coupled plasma optical emission spectroscopy with vacuum ultraviolet wavelength coverage and charge injection device detection, *Spectrochim. Acta, Part B*, 2001, v. 56, p. 1143-1157.

75. H. Becker-Ross, S. Florek, H. Franken, B. Radziuk, M. Zeiher, A scanning echelle monochromator for ICP-OES with dynamic wavelength stabilization and CCD detection, *J. Anal. At. Spectrom.*, 2000, v. 15, p. 851-861.

76. S. Florek, C. Haisch, M. Okruss, H. Becker-Ross, A new, versatile echelle spectrometer relevant to laser induced plasma applications, *Spectrochim. Acta, Part B*, 2001, v. 56, p. 1027-1034.

77. J.M. Mermet, Trends in instrumentation and data processing in ICP-AES, *J. Anal. At. Spectrom.*, 2002, v. 17, p. 1065-1071.

78. B. Rosenkranz, J. Bettmer, Microwave-induced plasma-optical emission spectrometry - fundamental aspects and applications in metal speciation analysis, *TrAC, Trends Anal. Chem.*, 2000, v. 19, p. 138-156.

79. Y. Yang, H.Q. Zhang, A.M. Yu, Q.H. Jin, Microwave plasma torch analytical atomic spectrometry, *Microchem. J.*, 2000, v. 66, p. 147-170.

80. S.Y. Lu, C.W. LeBlanc, M.W. Blades, Analyte ionization in the furnace atomization plasma excitation spectrometry source - spatial and temporal observations, *J. Anal. At. Spectrom.*, 2001, v. 16, p. 256-262.

81. P.Grinberg, R.C. Campos, Z. Mester, R.E. Sturgeon, A comparison of alkyl derivatization methods for speciation of mercury based on solid phase microextraction gas chromatography with furnace atomization plasma emission spectrometry detection, *J. Anal. At. Spectrom.*, 2003, v. 18, p. 902-909.

82. P.Grinberg, R.C. Campos, R.E. Sturgeon, Iridium as a permanent modifier for determination of cadmium and lead in sediment and biological samples by furnace atomization plasma emission spectrometry, *J. Anal. At. Spectrom.*, 2002, v. 17, p. 693-698.

83. J. Franzke, K. Kunze, M. Miclea, K. Niemax, Microplasmas for analytical spectrometry, *J. Anal. At. Spectrom.,* 2003, v. 18, p. 802-807.

84. V. Karanassios, Microplasmas for chemical analysis: analytical tools or research toys?, *Spectrochim. Acta, Part B*, 2004, v. 59, p. 909-928.

85. J.A.C. Broekaert, The development of microplasmas for spectrochemical analysis, *Anal. Bioanal. Chem.*, 2002, v. 374, p. 182-187.

86. E. Tognoni, V. Palleschi, M. Corsi, G. Cristoforetti, Quantitative micro-analysis by laser-induced breakdown spectroscopy: a review of the experimental approaches, *Spectrochim. Acta, Part B*, 2002, v. 57, p. 1115-1130.

87. R.E. Russo, X. Mao, H. Liu, J. Gonzalez, S.S. Mao, Laser ablation in analytical chemistry – a review, *Talanta*, 2002, v. 57, p. 423-607.

88. W.B. Lee, J.Y. Wu, Y.I. Lee, J. Sneddon, Recent applications of laser-induced breakdown spectrometry: A review of material approaches, *Appl. Spectrosc. Rev.*, 2004 , v. 39 , p. 27-97.

89. K. Song, Y.I. Lee, J. Sneddon, Recent developments in instrumentation for laser induced breakdown spectroscopy, *Appl. Spectrosc. Rev.*, 2002, v. 37, p. 89-117.

90. D. Günther, I. Horn, B. Hattendorf, Recent trends and developments in laser ablation-ICP-mass spectrometry, *Fresenius J. Anal. Chem.*, 2000, v. 368, p. 4-14.

91. R.E. Russo, X. Mao, J.J. Gonzalez, S.S. Mao, Femtosecond laser ablation ICP-MS, *J. Anal. At. Spectrom.*, 2002, v. 17, p. 1072-1075.

92. V. Sturm, R. Noll, Laser-induced breakdown spectroscopy of gas mixtures of air, $CO_2$, $N_2$, and $C_3H_8$ for simultaneous C, H, O, and N measurement, *Appl. Opt.*, 2003, v. 42, p. 6221-6225.

93. J.D. Hybl, G.A. Lithgow, S.G. Buckley, Laser-Induced Breakdown Spectroscopy Detection and Classification of Biological Aerosols, *Appl. Spectrosc.*, 2003, v. 57, p. 1207-1215.

94. J.E. Carranza, B.T. Fisher, G.D. Yoder, D.W. Hahn, On-line analysis of ambient air aerosols using laser-induced breakdown spectroscopy. *Spectrochim. Acta, Part B*, 2001, v. 56, p. 851‑864.

95. U. Panne, R. E. Neuhauser, C. Haisch, H. Fink, R. Niessner, Remote analysis of a mineral melt by laser-induced plasma spectroscopy, *Appl. Spectrosc.*, 2002, v. 56, p. 375-380.

96. L. Peter, V. Sturm, R. Noll, Liquid steel analysis with laser-induced breakdown spectrometry in the vacuum ultraviolet, *Appl. Opt.*, 2003, v. 42, p. 6199-6204.

97. R.G. Wiens, R.E. Arvidson, D.A. Cremers, M.J. Ferris, J.D. Blacic, F.P. Seelos, K.S. Deal, Combined remote mineralogical and elemental identification from rovers: field and laboratory tests using reflectance and laser-induced breakdown spectroscopy, *J. Geophys. Res., E (Planets)*, 2002, v. 107, p. FIDO3/1- 14.

98. A.K. Knight, N.L. Scherbarth, D.A. Cremers, M.J. Ferris, Characterization of laser-induced breakdown spectroscopy (LIBS) for application to space exploration. *Appl. Spectrosc.*, 2000, v. 54, p. 331-340.

99. J.J. Zayhowski, A.L. Wilson, Miniature, pulsed Ti:Sapphire laser system, *IEEE J. Quantum Electron.*, 2002, v. 38, p. 1449-1454.

100. В.И. Царёв, О.А. Букин, С.С. Голик, А.А. Ильин, М.В. Тренина, Применение метода лазерной искровой спектроскопии в экологических исследованиях, *Электрон. ж. "Исследовано в России"*, 2003, т. 6, с. 2108-2116 (http://zhurnal.ape.relarn.ru/articles/2003/174.pdf).

101. S. Palanco, A. Alises, J. Cunat, J. Baena, J.J. Laserna, Development of a portable laser-induced plasma spectrometer with fully-automated operation and quantitative analysis capabilities, *J. Anal. At. Spectrom.*, 2003, v. 18, p. 933-938.

102. M.D. Cheng, Real-time measurement of trace metals on fine particles by laser-induced plasma techniques, *Fuel Process. Technol.*, 2000, v. 65, p. 219-229.

103. I.B. Gornushkin, K. Amponsah-Manager, B.W. Smith, N. Omenetto, J.D. Winefordner, Microchip Laser-Induced Breakdown Spectroscopy: A Preliminary Feasibility Investigation, *Appl. Spectrosc.*, 2004, v. 58, p. 762-769.

104. А.А. Антипов, А.Э. Грасюк, С.В. Ефимовский, С.В. Курбасов, Л.Л. Лосев, В.И. Сосков, Повышение температуры лазерной плазмы при двухчастотном УФ-ИК воздействии на металлические мишени, *Квант. электрон.*, 1998, т. 25, с. 31-35.

105. L. St-Onge, V. Detalle, M. Sabsabi, Enhanced laser-induced breakdown spectroscopy using the combination of fourth-harmonic and fundamental Nd:YAG laser pulses, *Spectrochim. Acta, Part B*, 2002, v. 57, p. 121-135.

106. S.L. Lui, N.H. Cheung, Resonance-enhanced laser-induced plasma spectroscopy for sensitive elemental analysis: Elucidation of enhancement mechanisms, *Appl. Phys. Let.*, 2002, v. 81, p. 5114-5116.

107. Л.П. Житенко, Ю.В. Талдонов, Л.Е. Озерова, В.П. Обрезумов, Определение примесей в золоте методом атомно-эмиссионной спектрометрии с индуктивно-связанной плазмой и искровой абляцией, *Завод. лабор.*, 2001, т. 67, с. 22-24.

108. A.J.R. Hunter, R.T. Wainner, L.G. Piper, S.J. Davis, Rapid field screening of soils for heavy metals with spark-induced breakdown spectroscopy, *Appl. Opt.*, 2003, v. 42, p. 2102-2109.

109. A.J.R. Hunter, S.J. Davis, L.G. Piper, K.W. Holtzclaw, M.E.Fraser, Spark-induced breakdown spectroscopy: a new technique for monitoring heavy metals, *Appl. Spectrosc.*, 2000, v. 54, p. 575-582.

110. K. Wagatsuma, Emission characteristics of mixed gas plasmas in low-pressure glow discharges, *Spectrochim. Acta, Part B*, v. 56, p. 465-564.

111. K. Shimizu, H. Habazaki, P. Skeldon, G.E. Thompson, Impact of RF-GD-OES in practical surface analysis, *Spectrochim. Acta, Part B*, 2003, v. 58, p. 1573-1583.

112. S. Baude, J.A.C. Broekaert, D. Delfosse, N. Jakubowski, L. Fuechtjohann, N.G. Orellana-Velado, R. Pereiro, A. Sanz-Medel, Glow discharge atomic spectrometry for the analysis of environmental samples – a review, *J. Anal. At. Spectrom.*, 2000, v. 15, p. 1516-1525.

113. M. Betti, L.A. de las Heras, Glow discharge spectrometry for the characterization of nuclear and radioactivity contaminated environmental samples, *Spectrochim. Acta, Part B*, 2004, v. 59, p. 1359-1376.

114. В.С. Бураков, В.В. Кирис, П.А. Науменков, С.Н. Райков, Безэталонный лазерный спектральный анализ стёкол и медных сплавов, *Ж. прикл. спектроск.*, 2004, т. 71, с. 676-682.

115. D. Bulajic, M. Corsi, G. Cristoforetti, S. Legnaioli, V. Palleschi, A. Salvetti, E. Tognoni, A procedure for correcting self-absorption in calibration free–laser induced breakdown spectroscopy, *Spectrochim. Acta, Part B*, 2002, v. 57, p. 339-353.

116. A.A. Bol'shakov, G.G. Glavin, R.M. Barnes, Indirect semi-quantitative calibration of an enclosed ICP for the analysis of gases, *ICP Inf. Newslett.*, 2000, v. 26, p. 136-137.

117. J.S. Becker, H.-J. Dietze, State-of-the-art in inorganic mass spectrometry for analysis of high-purity materials, *Int. J. Mass Spectrom.*, 2003, v. 228, p. 127-150.

118. J.S. Becker, H.-J. Dietze, Inorganic mass spectrometric methods for trace, ultra-trace, isotope and surface analysis, *Int. J. Mass Spectrom.*, 2000, v. 195/196, p. 1-35.

119. G.M. Hieftje, J.H. Barnes, O.A. Gron, A.M. Leach, D.M. McClenathan, S.J. Ray, D.A. Solyom, W.C. Wetzel, M.B. Denton, D.W. Koppenaal, Evolution and revolution in instrumentation for plasma-source mass spectrometry, *Pure Appl. Chem.*, 2001, v. 73, p. 1579-1588.

120. J.S. Becker, Applications of inductively coupled plasma mass spectrometry and laser ablation inductively coupled plasma mass spectrometry in materials science, *Spectrochim. Acta, Part B*, 2002, v. 57, p. 1805-1820.

121. S.N. Ketkar, A.D. Scott, E.J. Hunter, The use of proton-transfer reactions to detect low levels of impurities in bulk oxygen using an atmospheric pressure ionization mass spectrometer, *Int. J. Mass Spectrom.*, 2001, v. 206, p. 7-12.

122. M. Kitano, Y. Shirai, A. Ohki, S. Babasaki, T. Ohmi, Impurity measurement in specialty gases using an atmospheric pressure ionization mass spectrometer with a two-compartment ion source, *Jpn. J. Appl. Phys., Part 1*, 2001, v. 40, p. 2688-2693.

123. K.P. Jochum, B. Stoll, J.A. Pfänder, M. Seufert, M. Flanz, P. Maissenbacher, M. Hofmann, A.W. Hofmann, Progress in multi-ion counting spark-source mass spectrometry (MIC-SSMS) for the analysis of geological samples, *Fresenius J. Anal. Chem.*, 2001, v. 370, p. 647-653.

124. N. Nonose, M. Kubota, Non-spectral and spectral interferences in inductively coupled plasma high-resolution mass spectrometry - Part 2. Comparison of interferences in quadrupole and high resolution inductively coupled plasma mass spectrometries, *J. Anal. At. Spectrom.*, 2001, v. 16, p. 560-566.

125. M. Moldovan, E.M. Krupp, A.E. Holliday, O.F.X. Donard, High resolution sector field ICP-MS and multicollector ICP-MS as tools for trace metal speciation in environmental studies: a review, *J. Anal. At. Spectrom.*, 2004, v. 19, p. 815 - 822.

126. Z. Cheng, Y. Zheng, R. Mortlock, A. van Geen, Rapid multi-element analysis of groundwater by high-resolution inductively coupled plasma mass spectrometry, *Anal. Bioanal. Chem.*, 2004, v. 379, p. 512-518.

127. J.M. Marchante-Gayon, Double-focusing ICP-MS for the analysis of biological materials, *Anal. Bioanal. Chem.*, 2004, v. 379, p. 335-337.

128. D.R. Bandura, V.I. Baranov, S.D. Tanner, Reaction chemistry and collisional processes in multipole devices for resolving isobaric interferences in ICP-MS, *Fresenius J. Anal. Chem.*, 2001, v. 370, p. 454-470.

129. D.W. Koppenaal, G.C. Eiden, C.J. Barinaga, Collision and reaction cells in atomic mass spectrometry: development, status, and applications, *J. Anal. At. Spectrom.*, 2004, v. 19, p. 561-570.

130. R.K. Marcus, Collisional dissociation in plasma source mass spectrometry: A potential alternative to chemical reactions for isobar removal, *J. Anal. At. Spectrom.*, 2004, v. 19, p. 591-599.

131. S.D. Tanner, V.I. Baranov, D. Bandura, Reaction cells and collision cells for ICP-MS: a tutorial review, *Spectrochim. Acta, Part B*, 2002, v. 57, p. 1361-1452.

132. D.A. Solyom, G.M. Hieftje, Simultaneous or scanning data acquisition? A theoretical comparison relevant to inductively coupled plasma sector-field mass spectrometers, *J. Am. Soc. Mass Spectrom.*, 2001, v. 14, p. 227-235.

133. D.A. Solyom, T.W. Burgoyne, G.M. Hieftje, Plasma-source sector mass spectrometry with array detection, *J. Anal. At. Spectrom.*, 1999, v.14, p. 1101-1110.

134. E. Engström, A. Stenberg, D.C. Baxter, D. Malinovsky, I. Mäkinen, S. Pönni, I. Rodushkin, Effects of sample preparation and calibration strategy on accuracy and precision in the multi-elemental analysis of soil by sector-field ICP-MS, *J. Anal. At. Spectrom.*, 2004, v. 19, p. 858-866.

135. S.J. Ray, G.M. Hieftje, Mass analyzers for inductively coupled plasma time-of-flight mass spectrometry, *J. Anal. At. Spectrom.*, 2001, v. 16, p. 1206-1216.

136. M. Balcerzak, An overview of analytical applications of time of flight-mass spectrometric (TOF-MS) analyzers and an inductively coupled plasma-TOF-MS technique, *Anal. Sci.*, 2003, v. 19, p. 979-989.

137. C.S. Westphal, J.A. McLean, B.W. Acon, L.A. Allen, A. Montaser, Axial inductively coupled plasma time-of-flight mass spectrometry using direct liquid sample introduction, *J. Anal. At. Spectrom.*, 2002, v. 17, p. 669-675.

138. D.M. McClenathan, S.J. Ray, W.C. Wetzel, G.M. Hieftje, Plasma source TOFMS, *Anal. Chem.*, 2004, v. 76, p. 158A-166A.

139. B. Hattendorf, C. Latkoczy, D. Gunther, Laser ablation ICPMS, *Anal. Chem.*, 2003, v. 75, p. 341A-347A.

140. J.H. Barnes, G.D. Schilling, G.M. Hieftje, R.P. Sperline, M.B. Denton, C.J. Barinaga, D.W. Koppenaal, Use of a novel array detector for the direct analysis of solid samples by laser ablation inductively coupled plasma sector-field mass spectrometry, *J. Am. Soc. Mass Spectrom.*, 2004, v. 15, p. 769-776.

141. A.M. Leach, G.M. Hieftje, Identification of alloys using single shot laser ablation inductively coupled plasma time-of-flight mass spectrometry, *J. Anal. At. Spectrom.*, 2002, v. 17, p. 852 - 857.

142. В.С. Севастьянов, Г.И. Рамендик, С.М. Сильнов, Н.Е. Бабулевич, Д.А. Тюрин, Е.В. Фатюшина, Влияние газовой среды в ионном источнике лазерного масс-спектрометра на результаты элементного анализа, *Ж. аналит. хим.*, 2002, т. 57, с. 509-515.

143. А.А. Сысоев, С.С. Потешин, А.Ю. Адамов, И.В. Мартынова, Аналитические возможности лазерного времяпролётного масс-спектрометра, *Завод. лабор.*, 2001, т. 67, с. 16-22.

144. A.A. Sysoev, A.A. Sysoev, Can laser-ionisation time-of-flight mass spectrometer be a promising alternative to laser ablation/inductively-coupled plasma mass spectrometry and glow discharge mass spectrometry for the elemental analysis of solids, *Europ. J. Mass Spectrom.*, 2002, v. 8, p. 213-232.

145. А.А. Сысоев, С.С. Потешин, Г.Б. Кузнецов, И.А. Ковалёв, Е.С. Юшков, Анализ компактных и порошковых проб с помощью лазерного времяпролётного масс-спектрометра ЛАМАС-10М, *Ж. аналит. хим.*, 2002, т. 57, с. 958-970.

146. В.В. Безруков, И.Д. Ковалёв, К.Н. Малышев, Д.К. Овчинников, Влияние ионного состава лазерной плазмы на метрологические характеристики результатов определений на тандемном лазерном масс-рефлектроне, *Ж. аналит. хим.*, 2004, т. 59, с. 723-728.

147. U. Rohner, J.A. Whitby, P. Wurz, A miniature laser ablation time-of-flight mass spectrometer for in situ planetary exploration, *Measur. Sci. Technol.*, 2003, v. 14, p. 2159-2164.

148. M. Yorozu, T. Yanagida, T. Nakajyo, Y. Okada, A. Endo, Laser microprobe and resonant laser ablation for depth profile measurements of hydrogen isotope atoms contained in graphite, *Appl. Opt.*, 2001, v. 40, p. 2043-2046.

149. K. Watanabe, K. Hattori, J. Kawarabayashi, T. Iguchi, Improvement of resonant laser ablation mass spectrometry using high-repetition-rate and short-pulse tunable laser system, *Spectrochim. Acta, Part B*, 2003, v. 58, p. 1163-1169.

150. D.M. Grim, J. Allison, Identification of colorants as used in watercolor and oil paintings by UV laser desorption mass spectrometry, *Int. J. Mass Spectrom.*, 2003, v. 222, p. 85-99.

151. M.V. Peláez, J.M. Costa-Fernández, R. Pereiro, N. Bordel, A. Sanz-Medel, Quantitative depth profile analysis by direct current glow discharge time of flight mass spectrometry, *J. Anal. At. Spectrom.*, 2003, v. 18, p. 864-871.

152. J. Pisonero, J.M. Costa, R. Pereiro, N. Bordel, A. Sanz-Medel, A radiofrequency glow-discharge-time-of-flight mass spectrometer for direct analysis of glasses, *Anal. Bioanal. Chem.*, 2004 , v. 379, p. 658-667.

153. A.A. Ganeev, M. Kuzmenkov, M.N. Slyadnev, M.V. Voronov, S.V. Potapov, Pulsed glow discharge with high flow rate of buffer gas and positive DC bias potential as ion source for TOFMS elemental analysis, *The Pittsburgh Conference on Analytical Chemistry and Applied Spectroscopy, Pittcon 2005, Orlando (USA, FL)*, Abstracts, 2005.

154. A. Chatterjee, Y. Shibata, A. Tanaka, M. Morita, Determination of selenoethionine by flow injection-hydride generation-atomic absorption spectrometry/high-performance liquid chromatography-hydride generation-high power nitrogen microwave-induced plasma mass spectrometry, *Anal. Chim. Acta*, 2001 , v. 436, p. 253-263.

155. A. Chatterjee, Y. Shibata, J. Yoshinaga, Tanaka, M. Morita, Determination of arsenic compounds by high performance liquid chromatography-ultrasonic nebulizer high power nitrogen microwave induced plasma mass spectrometry: an accepted coupling, *Anal. Chem.*, 2000, v. 72, p. 4402-4412.

156. I.I. Stewart, R.E. Sturgeon, Use of Ar-He mixed gas plasmas for furnace atomisation plasma ionisation mass spectrometry (FAPIMS), *J. Anal. At. Spectrom.*, 2000, v. 15, p. 1223-1232.

157. K.H. Rubin, Analysis of $^{232}$Th/$^{230}$Th in volcanic rocks: a comparison of thermal ionization mass spectrometry and other methodologies, *Chem. Geol.*, 2001, v. 175, p. 723-750.

158. L. Pajo, G. Tamborini, G. Rasmussen, K. Mayer, L. Koch, A novel isotope analysis of oxygen in uranium oxides: comparison of secondary ion mass spectrometry, glow discharge mass spectrometry and thermal ionization mass spectrometry, *Spectrochim. Acta, Part B*, 2001, v. 56, p. 541-549.

159. P. Ghosh, W.A. Brand, Stable isotope ratio mass spectrometry in global climate change research, *Int. J. Mass Spectrom.*, 2003, v. 228, p. 1-33.

160. L.R. Karam, L. Pibida, C.A. McMahon, Use of resonance ionization mass spectrometry for determination of Cs ratios in solid samples, *Appl. Radiat. Isot.*, 2002, v. 56, p. 369-374.

161. S.S. Dimov, S.L. Chryssoulis, R.H. Lipson, Quantitative elemental analysis for rhodium and palladium in minerals by time-of-flight resonance ionization mass spectrometry, *Anal. Chem.*, 2003, v. 75, p. 6723-6727.

162. M.R. Savina, M.J. Pellin, C.E. Tripa, I.V. Veryovkin, W.F. Calaway, A.M. Davis, Analyzing individual presolar grains with CHARISMA, *Geochem. Cosmochem. Acta*, 2003, v. 17, p. 3215-3225.

163. K. Blaum, C. Geppert, W.G. Schreiber, J.G. Hengstler, P. Müller, W. Nörtershäuser, K. Wendt, B.A. Bushaw, Trace determination of gadolinium in biomedical samples by diode laser-based multi-step resonance ionization mass spectrometry, *Anal. Bioanal. Chem.*, 2002, v. 372, p. 759-765.

164. M.J. Pellin, W.F. Calaway, A.M. Davis, R.S. Lewis, S. Amari, R.N. Clayton, Toward complete isotopic analysis of individual presolar silicon carbide grains: C, N, Si, Sr, Zr, Mo, and Ba in single grains of type X (abstract), *Lunar Planet. Sci.*, 2000, v. 31, p. 1917.

165. U. Köster, V.N. Fedoseyev, V.I. Mishin, Resonant laser ionization of radioactive atoms, *Spectrochim. Acta, Part B*, 2003, v. 58, p. 1047-1068.

166. K.D.A. Wendt, The new generation of resonant laser ionization mass spectrometers: becoming competitive for selective atomic ultra-trace determination?, *Europ. J. Mass Spectrom.*, 2002, v. 8, p. 273-285.

167. N. Trautmann, G. Passler, K.D.A. Wendt, Ultratrace analysis and isotope ratio measurements of long-lived radioisotopes by resonance ionization mass spectrometry (RIMS), *Anal. Bioanal. Chem.*, 2004, v. 378, p. 348-355.

168. M. Suter, 25 years of AMS – a review of recent developments, *Nucl. Instrum. Meth. Phys. Res., Sect. B*, 2004, v. 223/224, p. 139-148.

169. R. Hellborg, M. Faarinen, M. Kiisk, C. E. Magnusson, P. Persson, G. Skog, K. Stenström, Accelerator mass spectrometry – an overview, *Vacuum*, 2003, v. 70, p. 365-372.

170. L.K. Fifield, Applications of accelerator mass spectrometry: advances and innovation, *Nucl. Instrum. Meth. Phys. Res., Sect. B*, 2004, v. 223/224, p. 401-411.

171. P. Muzikar, D. Elmore, D.E. Granger, Accelerator mass spectrometry in geologic research, *Geolog. Soc. Amer. Bulletin*, 2003, v.115, p. 643-654.

172. C. Tuniz, U. Zoppi, M.A.C. Hotchkis, Sherlock Holmes counts the atoms, *Nucl. Instrum. Meth. Phys. Res., Sect. B*, 2004, v. 213, p. 469-475.

173. F. Li, Z. Xie, H. Schmidt, S. Sielemann, J.I. Baumbach, Ion mobility spectrometer for online monitoring of trace compounds, *Spectrochim. Acta, Part B*, 2002, v. 57, p. 1563-1574.

174. R.G. Ewing, D.A. Atkinson, G.A. Eiceman, G.J. Ewing, A critical review of ion mobility spectrometry for the detection of explosives and explosive related compounds, *Talanta*, 2001, v. 54, p. 515-529.

175. S.N. Ketkar, S. Dheandhanoo, Use of ion mobility spectrometry to determine low levels of impurities in gases, *Anal. Chem.*, 2001, v. 73, p. 2554-2557.

176. R.G. Longwitz, H. van Lintel, P. Renaud, Study of micro-glow discharges as ion sources for ion mobility spectrometry, *J. Vac. Sci. Technol. B*, 2003, v. 21, p. 1570-1573.

177. И.А. Буряков, Ю.Н. Коломиец, В.Б. Луппу, Обнаружение паров взрывчатых веществ в воздухе с помощью спектрометра нелинейности дрейфа ионов, *Ж. аналит. хим.*, 2001, т. 56, с. 381-386.

178. R. Clough, J. Truscatt, S.T. Belt, E.H. Evans, B. Fairman, T. Catterick, Isotope dilution ICP-MS for speciation studies, *Appl. Spectrosc. Rev.*, 2003, v. 38, p. 101-132.

179. K.G. Heumann, Isotope-dilution ICP-MS for trace element determination and speciation: from a reference method to a routine method?, *Anal. Bioanal. Chem.*, 2004, v. 378, p. 318-329.

180. А.А. Пупышев, А.К. Луцак, Возможности термодинамического моделирования термохимических процессов в плазме индуктивно связанного разряда, *Аналит. контроль*, 2000, т. 4, с. 304-315.

181. Г.И. Рамендик, Е.В. Фатюшина, А.И. Степанов, В.С. Севастьянов, Новый подход к расчёту коэффициентов относительной чувствительности в масс-спектрометрии с индуктивно связанной плазмой, *Ж. аналит. хим.*, 2001, т. 56, с. 566-574.

182. Г.И. Рамендик, Е.В. Фатюшина, А.И. Степанов, Обзорный масс-спектрометрический анализ с индуктивно связанной плазмой без использования стандартных образцов состава, *Ж. аналит. хим.*, 2003, т. 58, с. 20-27.

183. Г.И. Рамендик, Е.В. Фатюшина, А.И. Степанов, Определение неодима и иттербия методом лазерной масс-спектрометрии с изотопным разбавлением, *Ж. аналит. хим.*, 2003, т. 58, с. 174-178.

184. Н.А. Клюев, Применение масс-спектрометрии и хромато-масс-спектрометрии в анализе лекарственных препаратов, *Ж. аналит. хим.*, 2002, т. 57, с. 566-584.

185. В.А. Хвостиков, Г.Ф. Телегин, С.С. Гражулене, Определение ртути методом бездисперсионного атомно-флуоресцентного анализа, *Ж. аналит. хим.*, 2003, т. 58, с. 581-586.

186. H. Bagheri, A. Gholami, Determination of very low levels of dissolved mercury(II) and methylmercury in river waters by continuous flow with on-line UV decomposition and cold-vapor atomic fluorescence spectrometry after pre-concentration on a silica gel-2-mercaptobenzimidazol sorbent., *Talanta*, 2001, v. 55, p. 1141-1150.

187. L. Liang, M. Horvat, H. Li, P. Pang, Determination of mercury in minerals by combustion/trap/ atomic fluorescence spectrometry, *J. Anal. At. Spectrom.*, 2003, v. 18, p. 1383-1385.

188. A. D'Ulivo, E. Bramanti, L. Lampugnani, R. Zamboni, Improving the analytical performance of hydride generation non-dispersive atomic fluorescence spectrometry. Combined effect of additives and optical filters, *Spectrochim. Acta, Part B*, 2001, v. 56, p. 1893-1907.

189. N.V. Semenova, L.O. Leal, R. Forteza, V. Cerdà, Multisyringe flow injection system for total inorganic selenium determination by hydride generation–atomic fluorescence spectrometry, *Anal. Chim. Acta*, 2003, v. 486, p. 217-225.

190. A. Sayago, R. Beltrán, M.A.F. Recamales, J.L. Gómez-Ariza, Optimization of an HPLC-HG-AFS method for screening Sb(v), Sb(III), and $Me_3SbBr_2$ in water samples, *J. Anal. At. Spectrom.*, 2002, v. 17, p. 1440-1404.

191. S. Jianbo, T. Zhiyong, T. Chunchua, C. Quan, J. Zexiang, Determination of trace amounts of germanium by flow injection hydride generation atomic fluorescence spectrometry with on-line coprecipitation, *Talanta*, 2002, v. 56, p. 711-716.

192. H. Sun, R. Suo, Y. Lu, Determination of zinc in food using atomic fluorescence spectrometry by hydride generation from organized media, *Anal. Chim. Acta*, 2002, v. 457, p. 305-310.

193. J.T. Van Elteren, Z. Slejkovec, H.A. Das, A mathematical model for optimization of flow injection–hydride generation atomic fluorescence spectrometry, *Spectrochim. Acta, Part B*, 1999, v. 54, p. 311-320.

194. Aurora AI-3200: www.aurora-instr.com/AI3200NF.htm

195. A. Young, L. Pitts, S. Greenfield, M. Foulkes, A preliminary comparison of radial and axial excitation fluorescence in the ICP using non-laser sources, *J. Anal. At. Spectrom.*, 2003, v. 18, p. 44-48.

196. P. Stchur, K.X. Yang, X. Hou, T. Sun, R.G. Michel, Laser excited atomic fluorescence spectrometry – a review, *Spectrochim. Acta, Part B*, 2001, v. 56, p. 1565-1592.

197. Y. He, B.J. Orr, Tunable single-mode operation of a pulsed optical parametric oscillator pumped by a multimode laser, *Appl. Opt.*, 2001, v. 40, p. 4836-4848.

198. J. Mes, M. Leblans, W. Hogervorst, Single-longitudinal-mode optical parametric oscillator for spectroscopic applications, *Opt. Lett.*, 2002, v. 27, p. 1442-1444.

199. G.W. Baxter, M.A. Payne, B.D.W. Austin, C.A. Halloway, J.G. Haub, Y. He, A.P. Milce, J.F. Nibler, B.J. Orr, Spectroscopic diagnostics of chemical processes: applications of tunable optical parametric oscillators, *Appl. Phys. B*, 2000, v. 71, p. 651-663.

200. N.P. Barnes, Optical parametric oscillators, Chap. 7 in *Tunable Lasers Handbook*, F.J. Duarte, editor, Academic Press, San Diego (USA, CA), 1995, p. 293-348.

201. L. Burel, P. Giamarchi, L. Stephan, Y. Lijour, A. Le Bihan, Molecular and atomic ultra trace analysis by laser induced fluorescence with OPO system and ICCD camera, *Talanta*, 2003, v. 60, p. 295-302.

202. J.X. Zhou, X. Hou, K.X. Kang, R.G. Michel, Laser excited atomic fluorescence spectrometry in a graphite furnace with an optical parametric oscillator laser for sequential multi-element determination of cadmium, cobalt, lead, manganese and thallium in Buffalo river sediment, *J. Anal. At. Spectrom.*, 1998, v. 13, p. 41-47.

203. S. Aubin, E. Gomez, L.A. Orozco, G.D. Sprouse, High efficiency magneto-optical trap for unstable isotopes, *Rev. Sci. Instrum.*, 2003, v. 74, p. 4342-4351.

204. K. Bailey, C.Y. Chen, X. Du, Y.M. Li, Z.T. Lu, T.P. O'Connor, L. Young, ATTA - A new method of ultrasensitive isotope trace analysis, *Nucl. Instrum. Meth. Phys. Res., Sect. B*, 2000, v. 172, p. 224-227.

205. I.D. Moore, K. Bailey, J. Greene, Z.T. Lu, P. Müller, T.P. O'Connor, C. Geppert, K.D.A. Wendt, L. Young, Counting individual $^{41}$Ca atoms with a magneto-optical trap, *Phys. Rev. Lett.*, 2004, v. 92, p. 153002/1-4.

206. O.M. Marago, B. Fazio, P.G. Gucciardi, E. Arimondo, Atomic gallium laser spectroscopy with violet/blue diode lasers, *Appl. Phys. B*, 2003, v. 77, p. 809-815.

207. P.E. Walters, T.E. Barber, M.W. Wensing, J.D. Winefordner, A diode laser wavelength reference system applied to the determination of rubidium in atomic fluorescence spectroscopy, *Spectrochim. Acta, Part B*, 1991, v. 46, p. 1015-1020.

208. A. Zybin, C. Schnürer-Patschan, K. Niemax, Simultaneous multielement determination in a commercial graphite furnace by diode laser induced fluorescence, *Spectrochim. Acta, Part B*, 1992, v. 47, p. 1519-1524.

209. K. Yuasa, Y. Yamashina, T. Sakurai, Ar lowest excited state densities in Ar and Ar-Hg hot cathode discharge, *Jpn. J. Appl. Phys.*, 1997, v. 36, p. 2340-2345.

210. C. Raab, J. Bolle, H. Oberst, J. Eschner, F. Schmidt-Kaler, R. Blatt, Diode laser spectrometer at 493 nm for single trapped Ba$^+$ ions, *Appl. Phys. B*, 1998, v. 67, p. 683-688.

211. G.D. Severn, D.A. Edrich, R. McWilliams, Argon ion laser-induced fluorescence with diode lasers, *Rev. Sci. Instrum.*, 1998, v. 69, p. 10-15.

212. B.W. Smith, A. Quentmeier, M. Bolshov, K. Niemax, Measurements of uranium isotope ratios in solid samples using laser ablation and diode laser-excited atomic fluorescence spectrometry, *Spectrochim. Acta, Part B*, 1999, v. 54, p. 943-958.

213. J.P. Temirov, O.I. Matveev, B.W. Smith, J.D. Winefordner, Laser-enhanced ionization with avalanche amplification: detection of cesium at fg/mL levels, *Appl. Spectrosc.*, 2003, v. 57, p. 729-732.

214. W.L. Clevenger, O.I. Matveev, S. Cabredo, N. Omenetto, B.W. Smith, J.D. Winefordner, Laser-enhanced ionization of mercury atoms in an inert atmosphere with avalanche amplification of the signal, *Anal. Chem.*, 1997, v. 69, p. 2232-2237.

215. W.L. Clevenger, L.S. Mordoh, O.I. Matveev, N. Omenetto, B.W. Smith, J. D. Winefordner, Analytical time-resolved laser enhanced ionization spectroscopy I. Collisional ionization and photoionization of the Hg Rydberg states in a low pressure gas, *Spectrochim. Acta, Part B*, 1997, v. 52, p. 295-304.

216. J.F. Gravel, P. Nobert, J.F.Y. Gravel, D. Boudreau, Trace level determination of lead in solid samples by UV laser ablation and laser-enhanced ionization detection, *Anal. Chem.*, 2003, v. 75, p. 1442-1449.

217. H.L. Pacquette, S.A. Elwood, M. Ezer, J.B. Simeonsson, A comparison of continuous flow hydride generation laser-induced fluorescence and laser-enhanced ionization spectrometry approaches for parts per trillion level measurements of arsenic, selenium and antimony, *J. Anal. At. Spectrom.*, 2001, v. 16, p. 152-158.

218. K.L. Riter, O.I. Matveev, B.W. Smith, J.D. Winefordner, The determination of lead in whole blood by laser enhanced ionization using a combination of electrothermal vaporizer and flame, *Anal. Chim. Acta*, 1996, v. 333, p. 187-192.

219. C.-B. Ke, K.-C. Lin, Laser-enhanced ionization detection of Pb in seawater by flow injection analysis with on-line preconcentration and separation, *Anal. Chem.*, 1999, v. 71, p. 1561-1567.

220. C.-B. Ke, K.-D. Su, K.-C. Lin, Laser-enhanced ionization and laser-induced atomic fluorescence as element-specific detection methods for gas chromatography. Application to organotin analysis, *J. Chromatogr. A*, 2001, v. 921. p. 247-253.

221. А.А. Горбатенко, Н.Б. Зоров, Т.А. Лабутин, Использование атомно-ионизационной спектроскопии с лазерным испарением в пламя для анализа труднолетучих проб, *Ж. аналит. хим.*, 2003, т. 58, с.388-391.

222. А.А. Горбатенко, Н.Б. Зоров, А.Р. Муртазин, Формирование сигнала в лазерной атомно-ионизационной спектрометрии при лазерном пробоотборе в пламя, *Ж. аналит. хим.*, 2002, т. 57, с. 151-158.

223. J.F.Y. Gravel, M.L. Viger, P. Nobert, D. Boudreau, Multielemental laser-enhanced ionization spectrometry for the determination of lead at the trace level in pelletized coal using laser ablation and internal standard signal normalization, *Appl. Spectrosc.*, 2004, v. 58, p. 727-733.

224. A.T. Khalmanov, D.-K. Ko, J. Lee, N. Eshkobilov, A. Tursunov, Study of traces of Au and Ag atoms by resonant laser stepwise ionization spectroscopy, *J. Korean Phys. Soc.*, 2004, v. 44, p. 843-848.

225. J.B. Simeonsson, S.A. Elwood, M. Ezer, H.L. Pacquette, D.J. Swart, H.D. Beach, D.J. Thomas, Development of ultratrace laser spectrometry techniques for measurements of arsenic, *Talanta*, 2002, v. 58, p. 189-199.

226. А.Т. Халманов, Х.С. Хамраев, А.Т. Турсунов, О. Тухлибоев, Исследование следов европия методом лазерной резонансной ионизационной спектрометрии, *Опт. спектроск.*, 2002, т. 92, с. 885-887.

227. А.Т. Халманов, Х.С. Хамраев, А.Т. Турсунов, О. Тухлибоев, Исследование следов элементов на универсальном лазерно-фотоионизационном спектрометре, *Опт. спектроск.*, 2001, т. 90, с. 400-403.

228. D. Boudreau, J.F. Gravel, Laser-enhanced ionization: recent developments, *Trends Anal. Chem.*, 2001, v. 20, p. 20-27.

229. J.C. Travis, G.C. Turk, Editors, *Laser-Enhanced Ionization Spectrometry*, J. Wiley & Sons (Ser.: Chem. Analysis, v. 136), New York, 1996, 334 p.

230. B. Barbieri, N. Beverini, A. Sasso, Optogalvanic spectroscopy, *Rev. Modern Phys.*, 1990, v. 62, p. 603-644.

231. X. Yao, S.P. McGlynn, R.C. Mohanty, Laser optogalvanic analysis in a radiofrequency plasma: detection of iodine atoms and molecules, *Microchem. J.*, 1999, v. 61, p. 223-239.

232. А.Т. Халманов, Лазерная оптогальваническая спектроскопия атома индия, *Ж. прикл. спектроск.*, 2000, т. 67, с. 396-398.

233. J.P. Young, R.W. Shaw, C.M. Barshick, J.M. Ramsey, Determination of actinide isotope ratios using glow discharge optogalvanic spectroscopy, *J. Alloys Comounds*, 1998, v. 271-273, p. 62-65.

234. E.C. Jung, T.-S. Kim, K. Song, C.-J. Kim, Diode laser-excited optogalvanic and absorption measurements of uranium in a hollow cathode discharge, *Spectrosc. Lett.*, 2003, v. 36, p. 167-180.

235. T.J. McCarthy, M. Salvermoser, D.E. Murnick, Low-temperature opto-galvanic spectroscopy without discharges, *Hyperfine Interact.*, 2003, v. 151/152, p. 209-222.

236. А.А. Ганеев, В.Н. Григорьян, А.И. Дробышев, М.Н. Шляднев, С.Е. Шолупов, Аналитическая резонансно-ионизационная спектрометрия с импульсной ионизацией образца в полом катоде, *Ж. аналит. хим.*, 1996, т. 51, с. 845-854.

237. D.L. Monts, Abhilasha; S. Qian; D. Kumar; X. Yao; S.P. McGlynn, Comparison of atomization sources for a field-deployable laser optogalvanic spectrometry system, *J. Thermophys. Heat Transfer*, 1998, v. 12, p. 66-72.

238. E. de Oliveira, Sample preparation for atomic spectroscopy: Evolution and future trends, *J. Braz. Chem. Soc.*, 2003, v. 14, p. 174-182.

239. J. Pawliszyn, Sample Preparation: Quo Vadis?, *Anal. Chem.*, 2003, v. 75, p. 2543-2558.

240. И.В. Кубракова, Микроволновое излучение в аналитической химии: возможности и перспективы использования, *Успехи хим.*, 2002, т. 71, с. 327-340.

241. J.A. Nóbrega, L.C. Trevizan, G.C.L. Araújo, A.R.A. Nogueira, Focused-microwave-assisted strategies for sample preparation, *Spectrochim. Acta, Part B*, 2002, v. 57, p. 1855-1876.

242. Ю.А. Золотов, Г.И. Цизин, Е.И. Моросанова, С.Г. Дмитриенко, Сорбционное концентрирование микрокомпонентов для целей химического анализа, *Успехи хим.*, 2005, т. 74, с. 41-66.

243. J.C.A. Wuilloud, R.G. Wuilloud, A.P. Vonderheide, J.A. Caruso, Gas chromatography/plasma spectrometry – an important analytical tool for elemental speciation studies, *Spectrochim. Acta, Part B*, 2004, v. 59, p. 755-792.

244. M. de G. Pereira, M.A.Z. Arruda, Trends in preconcentration procedures for metal determination using atomic spectrometry techniques, *Mikrochim. Acta*, 2003, v. 141, p. 115-131.

245. K.G.Heumann, Hyphenated techniques - the most commonly used method for trace elemental speciation analysis, *Anal. Bioanal. Chem*., 2002, v. 373, p. 323-324.

246. А.И. Сапрыкин, И.Р. Шелпакова, Т.А. Чанышева, Н.П. Заксас, Некоторые аспекты подготовки проб к атомно-эмиссионному спектральному и масс-спектрометрическому определению микроэлементов, *Ж. аналит. хим.*, 2003, т. 58, с. 273-280.

247. М.А. Большов, А.В. Зыбин, Д. Буптчер, К. Нимакс, Определение отдельных форм соединений комбинированным методом ЖХ с элементоселективным детектором ААС с диодным лазером, *Ж. аналит. хим.*, 2003, т. 58, с.725-725.

248. X.-P. Yan, Y. Jiang, Flow injection on-line preconcentration and separation coupled with atomic (mass) spectrometry for trace element (speciation) analysis based on sorption of organo-metallic complexes in a knotted reactor, *TrAC, Trends Anal. Chem.*, 2001, v. 20, p. 552-562.

249. J.H. Wang, E.H. Hansen, Sequential injection lab-on-valve: the third generation of flow injection analysis, *TrAC, Trends Anal. Chem.*, 2003, v. 22, p. 225-231.

250. R.C. Prados-Rosales, J.L. Luque-García, M.D.L. de Castro, Valves and flow injection manifolds: an excellent marriage with unlimited versatility, *Anal. Chem. Acta*, 2003, v. 480, p. 181-192.

251. J.H. Wang, E.H. Hansen, On-line sample-pre-treatment schemes for trace-level determinations of metals by coupling flow injection or sequential injection with ICP-MS, *TrAC, Trends Anal. Chem.*, 2003, v. 22, p. 836-846.

252. A. D'Ulivo, C. Baiocchi, E. Pitzalis, M. Onor, R. Zamboni, Chemical vapor generation for atomic spectrometry. A contribution to the comprehension of reaction mechanisms in the generation of volatile hydrides using borane complexes, *Spectrochim. Acta, Part B*, 2004, v. 59, p. 471-486.

253. Y.-L. Feng, R.E. Sturgeon, J.W. Lam, Chemical vapor generation characteristics of transition and noble metals reacting with tetrahydroborate(III). *J. Anal. At. Spectrom.*, 2003, v. 18, p. 1435-1442.

254. E. Bolea, F. Laborda, J.R. Castillo, R.E. Sturgeon, Electrochemical hydride generation for the simultaneous determination of hydride forming elements by inductively coupled plasma-atomic emission spectrometry, *Spectrochim. Acta, Part B*, 2004, v. 59, p. 505-513.

255. P. Smichowski, S. Farias, Advantages and analytical applications of chloride generation. A review on vapor generation methods in atomic spectrometry, *Microchem. J.*, 2000, v. 67, p. 147-155.

256. X.C. Duan, R.L. McLaughlin, I.D. Brindle, A. Conn, Investigations into the generation of Ag, Au, Cd, Co, Cu, Ni, Sn and Zn by vapour generation and their determination by inductively coupled plasma atomic emission spectrometry, together with a mass spectrometric study of volatile species. Determination of Ag, Au, Co, Cu, Ni and Zn in iron, *J. Anal. At. Spectrom.*, 2002, v. 17, p. 227-231.

257. M. Resano, M. Verstraete, F. Vanhaecke, L. Moens, Evaluation of the multi-element capabilities of electrothermal vaporization quadrupole-based ICP mass spectrometry, *J. Anal. At. Spectrom.*, 2001, v. 16, p. 1018-1027.

258. M.J. Cal-Prieto, M. Felipe-Sotelo, A. Carlosena, J.M. Andrade, P. Lopez-Mahia, S. Muniategui, D. Prada, Slurry sampling for direct analysis of solid materials by electrothermal atomic absorption spectrometry (ETAAS). A literature review from 1990 to 2000, *Talanta*, 2002, v. 56, p. 1-51.

259. T. Kántor, Electrothermal vaporization and laser ablation sample introduction for flame and plasma spectrometric analysis of solid and solution samples, *Spectrochim. Acta, Part B*, 2001, v. 56, p. 1523-1563.

260. A.S. Luna, H.B. Pereira, I. Takase, R.A. Goncalves, R.E. Sturgeon, R.C. de Campos, Chemical vapor generation – electrothermal atomic absorption spectrometry: new perspectives, *Spectrochim. Acta, Part B*, 2002, v. 57, p. 2047-2056.

261. J.L. Burguera, M. Burguera, Flow injection-electrothermal atomic absorption spectrometry configurations: recent developments and trends, *Spectrochim. Acta, Part B*, 2001, v. 56, p. 1801-1829.

262. X. Hou, B.T. Jones, Tungsten devices in analytical atomic spectrometry, *Spectrochim. Acta, Part B*, 2002, v. 57, p. 659-688.

263. J.A. Nóbrega, J. Rust, C.P. Calloway, B.T. Jones, Use of modifiers with metal atomizers in electrothermal AAS: a short review, *Spectrochim. Acta, Part B*, 2004, v. 59, p. 1337-1345.

264. M. Huang, H. Kojima, T. Shirasaki, A. Hirabayashi, H. Koizumi, Study on solvent-loading effect on inductively coupled plasma and microwave-induced plasma sources with a microliter nebulizer, *Anal. Chim. Acta*, 2000, v. 413, p. 217-222.

265. H. Matusiewicz, A novel sample introduction system for microwave-induced plasma optical emission spectrometry, *Spectrochim. Acta, Part B*, 2002, v. 57, p. 485-494.

266. A.L. Rosen, G.M. Hieftje, Inductively coupled plasma mass spectrometry and electrospray mass spectrometry for speciation analysis: applications and instrumentation, *Spectrochim. Acta, Part B*, 2004, v. 59, p. 135-146.

267. P.J. Dyson, A.K. Hearley, B.F.G. Johnson, J.S. McIndoe, P.R.R. Langridge-Smith, C. Whyte, Combining energy-dependent electrospray ionisation with tandem mass spectrometry for the analysis of inorganic compounds, *Rapid Comm. Mass Spectrom.*, 2001, v. 15, p. 895-897.

268. I.I. Stewart, Electrospray mass spectrometry: a tool for elemental speciation, *Spectrochim. Acta, Part B*, 1999, v. 54, p. 1649-1695.

269. P. Rychlovský, P. Černoch, M. Skleničková, Application of a heated electrospray interface for on-line connection of the AAS detector with HPLC for detection of organotin and organolead compounds, *Anal. Bioanal. Chem.*, 2002, v. 374, p. 955-962.

270. B. Lavine, J.J. Workman, Chemometrics, *Anal. Chem.*, 2004, v. 76, p. 3365-3372.

271. P. Geladi, Chemometrics in spectroscopy. Part 1. Classical chemometrics, *Spectrochim. Acta, Part B*, 2003, v. 58, p. 767-782.

272. P. Geladi, B. Sethson, J. Nyström, T. Lillhonga, T. Lestander, J. Burger, Chemometrics in spectroscopy. Part 2. Examples, *Spectrochim. Acta, Part B*, 2004, v. 59, p. 1347-1357.

273. C.M. Wehlburg, D.M. Haaland, D.K. Melgaard, New hybrid algorithm for transferring multivariate quantitative calibrations of intra-vendor near-infrared spectrometers. *Appl. Spectrosc.*, 2002, v. 56, p. 877-886.

274. H.S. Yang, P.R. Griffiths, J.D. Tate, Comparison of partial least squares regression and multi-layer neural networks for quantification of nonlinear systems and application to gas phase Fourier transform infrared spectra. *Anal. Chim. Acta*, 2003, v. 489, p. 125-136.

275. М.А. Проскурнин, М.Ю. Кононец, Современная аналитическая термооптическая спектроскопия, *Успехи хим.*, 2004, т. 73, с. 1235.

276. R.E. Russo, X.L. Mao, C. Liu, J. Gonzalez, Laser assisted plasma spectrochemistry: laser ablation, *J. Anal. At. Spectrom.*, 2004, v. 19, p. 1084-1089.

277. J.L. Todoli, J.M. Mermet, Elemental analysis of liquid microsamples through inductively coupled plasma spectrochemistry, *TrAC, Trends Anal. Chem.*, 2005, v. 24, p. 107-116.

278. J.A.C. Broekaert, V. Siemens, Recent trends in atomic spectrometry with microwave-induced plasmas, *Spectrochim. Acta, Part B*, 2004, v. 59, p. 1823-1839.

279. V. Hoffmann, M. Kasik, P.K. Robinson, C. Venzago, Glow discharge mass spectrometry, *Anal. Bioanal. Chem.*, 2005, v. 381, p. 173-188.